\documentclass[aps,prd,preprintnumbers,superscriptaddress,showpacs,showkeys,floatfix,amssymb,amsfonts,twocolumn]{revtex4}  
\usepackage{MnSymbol}
\usepackage{hyperref}
\usepackage[usenames]{color}
\usepackage{bbm}
\usepackage{slashed}
\usepackage{amsmath}
\usepackage{multirow}
\usepackage{graphicx}
\usepackage{grffile}
\usepackage{xfrac}
\newcommand{\Nfa}{$\textrm{N}_\textrm{f}{=}2$}
\newcommand{\Nfb}{$\textrm{N}_\textrm{f}{=}2{+}1{+}1$}
\newcommand{\Nfc}{$\textrm{N}_\textrm{f}{=}4$}
\begin{document}

\title{Moments of nucleon generalized parton distributions from
  lattice QCD simulations at physical pion mass
}

\author{C.~Alexandrou}
\affiliation{Department of Physics, University of Cyprus, P.O. Box 20537, 1678 Nicosia, Cyprus} \affiliation{Computation-based Science and Technology Research Center, The Cyprus Institute, 20 Kavafi Str., Nicosia 2121, Cyprus}

\author{S.~Bacchio}
\affiliation{Computation-based Science and Technology Research Center, The Cyprus Institute, 20 Kavafi Str., Nicosia 2121, Cyprus}

\author{M.~Constantinou}
\affiliation{Department of Physics, Temple University, 1925 N. 12th Street, Philadelphia, Pennsylvaniaip. 19122-1801, USA}

\author{P.~Dimopoulos}
\affiliation{Dipartimento di Fisica, Universit{\`a} and INFN di Roma Tor Vergata, 00133 Roma, Italy}

\author{J.~Finkenrath}
\affiliation{Computation-based Science and Technology Research Center, The Cyprus Institute, 20 Kavafi Str., Nicosia 2121, Cyprus} 

\author{R.~Frezzotti}
\affiliation{Dip. di Fisica, Universit{\`a} and INFN di Roma Tor Vergata, 00133 Roma, Italy}

\author{K.~Hadjiyiannakou}
\affiliation{Computation-based Science and Technology Research Center, The Cyprus Institute, 20 Kavafi Str., Nicosia 2121, Cyprus} 

\author{K.~Jansen}
\affiliation{NIC, DESY, Platanenallee 6, D-15738 Zeuthen, Germany}

\author{B.~Kostrzewa}
\affiliation{HISKP (Theory), Rheinische Friedrich-Wilhelms-Universit{\"a}t Bonn, Nu{\ss}allee 14-16,  53115 Bonn, Germany}

\author{G.~Koutsou}
\affiliation{Computation-based Science and Technology Research Center, The Cyprus Institute, 20 Kavafi Str., Nicosia 2121, Cyprus}

\author{C.~Lauer}
\affiliation{Department of Physics, Temple University, 1925 N. 12th Street, Philadelphia, PA 19122-1801, USA} \affiliation{Physics Division, Argonne National Laboratory, Argonne, Illinois, 60439, USA}

\author{C.~Urbach}
\affiliation{HISKP (Theory), Rheinische Friedrich-Wilhelms-Universit{\"a}t Bonn, Nu{\ss}allee 14-16,  53115 Bonn, Germany}

\collaboration{
  Extended Twisted Mass Collaboration
}

\date{\today}

\begin{abstract}
  \centerline{\includegraphics[width=0.2\linewidth]{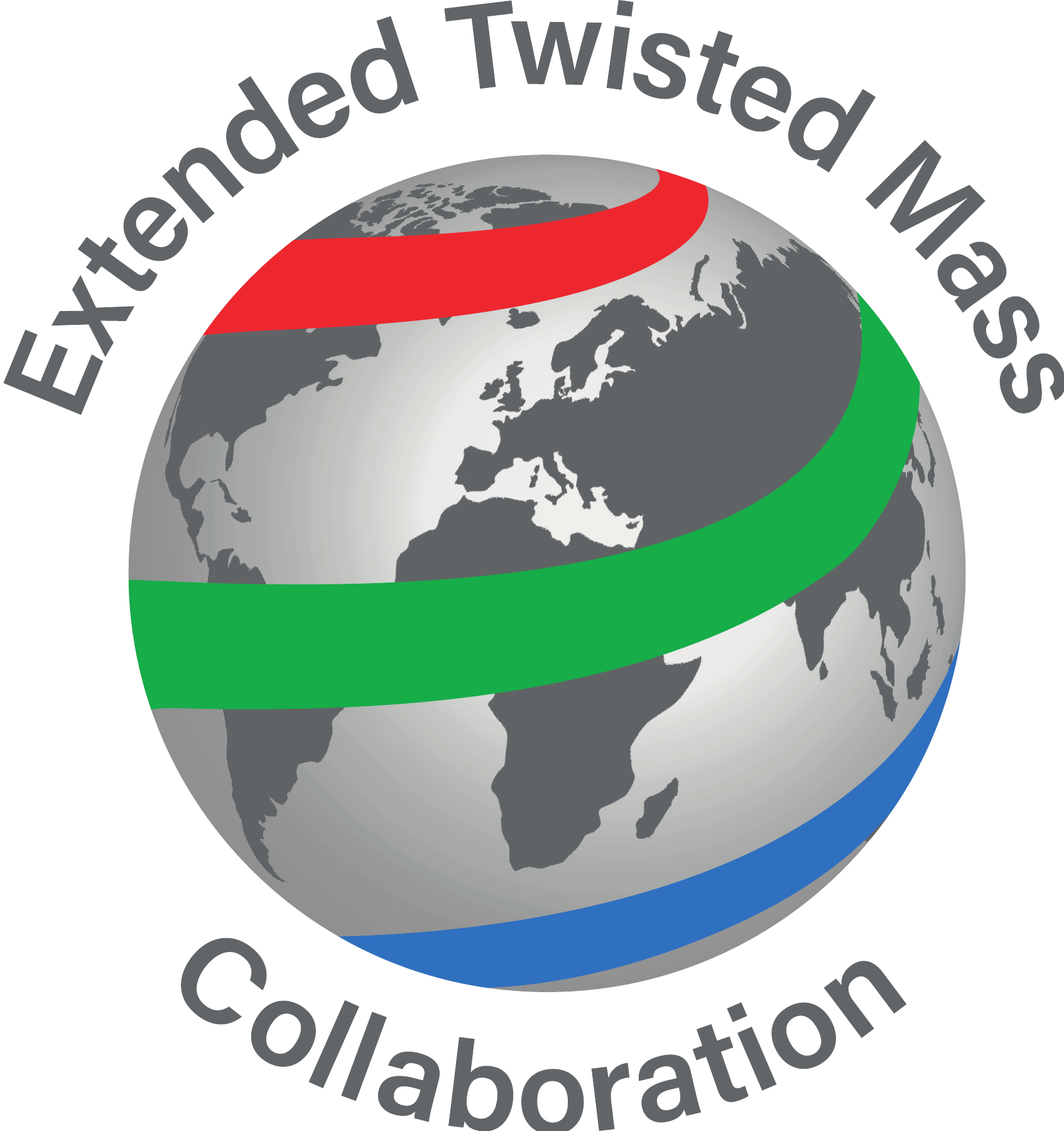}}

 We present results for the moments of nucleon isovector vector and
 axial generalized parton distribution functions computed within
 lattice QCD.  Three ensembles of maximally twisted mass
 clover-improved fermions simulated with a physical value of the pion
 mass are analyzed. Two of these ensembles are generated using two
 degenerate light quarks. A third ensemble is used having, in addition
 to the light quarks, strange and charm quarks in the sea.  A careful
 analysis of the convergence to the ground state is carried out that
 is shown to be essential for extracting the correct nucleon matrix
 elements.  This allows a controlled determination of the unpolarized,
 helicity and tensor second Mellin moments.  The vector and
 axial-vector generalized form factors are also computed as a function
 of the momentum transfer square up to about 1~GeV$^2$. The three
 ensembles allow us to check for unquenching effects and to assess
 lattice finite volume effects.

\end{abstract}
  
\maketitle

\section{Introduction}
Understanding the structure of the nucleon in terms of its fundamental
constituents is considered a milestone of hadronic physics. During the
past decades, parton distribution functions (PDFs) measured at
experimental facilities, such as HERA, RHIC, and LHC, have provided
valuable insights into the distribution of quarks and gluons within
the nucleon. Better determination of PDFs has also helped interpret
experimental data and provided input for ongoing and future
experiments. Furthermore, the planned Electron-Ion Collider envisions
a rich program of measurements, paving the way for nucleon tomography
and for mapping the three-dimensional structure of the nucleon.

Obtaining these quantities from first principles is one of the main
objectives of lattice QCD, which has seen remarkable progress in
recent years. In particular, the recent availability of simulations at
the physical values of the quark masses allows for obtaining nucleon
matrix elements without the need for a chiral extrapolation, thus
eliminating a major source of systematic error. In addition,
theoretical progress has enabled the first exploratory study of the
parton distribution functions themselves on the lattice as compared to
the traditional approach of calculating their
moments~\cite{Ji:2013dva}. While this is a promising approach,
progress still needs to be made in order to be able to have a direct
quantitative comparison with experiment. Therefore, the calculation of
moments on the lattice is crucial for comparing results with
experiment, especially as statistical precision for these quantities
increases and remaining systematic uncertainties, such as those from
the finite lattice spacing, from the finite volume, and from excited state
contaminations, come under control.

The generalized parton distributions (GPDs) occur in several physical
processes, such as deeply virtual Compton scattering and deeply
virtual meson production. Their forward limit coincides with the usual
parton distributions and their first moments are related to the
nucleon form factors. Since GPDs can be accessed in high energy
processes where QCD factorization applies, the amplitude can be
written as a convolution of a hard perturbative kernel and the
nonperturbative universal parton distributions.  GPDs are defined as
matrix elements of bilocal operators separated by a lightlike
interval.  A common approach is to proceed with an operator product
expansion that leads to a tower of local operators, the nucleon matrix
elements of which can be evaluated within lattice QCD.  In this paper,
we compute the nucleon matrix elements of the one-derivative operators
\begin{align}
  \mathcal{O}_{V}^{\mu\nu} =& \bar{\psi}\gamma^{\{\mu}\overleftrightarrow{D}^{\nu\}}\frac{\tau^3}{2}\psi,\nonumber\\
  \mathcal{O}_{A}^{\mu\nu} =& \bar{\psi}\gamma_5\gamma^{\{\mu}\overleftrightarrow{D}^{\nu\}}\frac{\tau^3}{2}\psi,\,\textrm{and}\nonumber\\
  \mathcal{O}_{T}^{\mu\nu\rho} =& \bar{\psi}\sigma^{[\mu\{\nu]}\overleftrightarrow{D}^{\rho\}}\frac{\tau^3}{2}\psi,\label{eq:operators}
\end{align}
where $\psi$ and $\bar{\psi}$ are light quark flavor doublets,
i.e. $\bar{\psi}=(\bar{u}, \bar{d})$. In this work, we consider isovector quantities, obtained using the
Pauli matrix $\tau^3$ as in Eq.~(\ref{eq:operators}).  The curly brackets denote
symmetrization and the square brackets antisymmetrization of the
enclosed indices, with subtraction of the trace implied whenever
symmetrizing and:
\begin{equation}
  \overleftrightarrow{D}_\mu = \frac{1}{2}(\overrightarrow{D}_\mu - \overleftarrow{D}_\mu),\, D_\mu = \frac{1}{2}(\nabla_\mu + \nabla_\mu^*)
\end{equation}
with $\nabla_\mu$ and $\nabla_\mu^*$ denoting the forward and backward
derivatives on the lattice, respectively. These nucleon matrix elements
can be expanded in terms of generalized form factors (GFFs), which are
Lorentz invariant functions of the momentum transfer squared. At zero
momentum transfer, these nucleon matrix elements yield the second Mellin
moments of the unpolarized, helicity and transversity PDFs.

In this paper we use three ensembles of twisted mass fermions with two
values of the lattice spacing and two physical volume sizes to compute
the three second Mellin moments. We also compute the GFFs related to
the vector and axial matrix elements. The parameters of the three
ensembles allow us to assess volume effects and check for any
indication of unquenching due to strange and charm quarks.

The remainder of this paper is organized as follows: in
Sec.~\ref{sec:matrix elements} we present the matrix elements used
and expressions for the GFFs obtained, in
Sec.~\ref{sec:methodology} we present the methodology employed for
extracting the GFFs from the lattice, details on the lattice ensembles
used, and parameters of our lattice analysis, with
Sec.~\ref{sec:renormalization} detailing the renormalization
procedure employed. In Sec.~\ref{sec:results} we provide our
results, and in Sec.~\ref{sec:conclusions} we give our conclusions.

\vspace*{0.5cm}

\section{Matrix elements}
\label{sec:matrix elements}

We consider the nucleon matrix elements $\langle
N(p',s')|\mathcal{O}^{\mu\nu}_\mathcal{H} |N(p,s)\rangle$ of the
three one-derivative operators of Eq.~(\ref{eq:operators}) where $s$
and $p$ ($s'$ and $p'$) are the initial (final) spin and momentum of
the nucleon, and $\mathcal{H}$ denotes the $\gamma$ structure
corresponding to the vector ($V$), axial ($A$) and tensor ($T$) operators.
 In the isovector
combination, the disconnected contributions cancel, leaving only
connected contributions. The nucleon matrix elements of the operators
of Eq.~(\ref{eq:operators}) can be written in terms of the generalized form factors
 as follows:

\begin{widetext}
  \begin{align}
    \langle N(p^\prime,s^\prime)| \mathcal{O}_{V}^{\mu\nu}| N(p,s)\rangle =
    \bar u_N(p^\prime,s^\prime)\frac{1}{2}\Bigl[& A_{20}(q^2)\, \gamma^{\{ \mu}P^{\nu\}} +B_{20}(q^2)\, \frac{i\sigma^{\{\mu\alpha}q_{\alpha}P^{\nu\}}}{2m_N}+C_{20}(q^2)\, \frac{1}{m_N}q^{\{\mu}q^{\nu\}} \Bigr] u_N(p,s), \nonumber \\
    \langle N(p^\prime,s^\prime)| \mathcal{O}_{A}^{\mu\nu}| N(p,s)\rangle =
    \bar u_N(p^\prime,s^\prime)\frac{i}{2}\Bigl[& \tilde A_{20}(q^2)\, \gamma^{\{\mu}P^{\nu\}}\gamma^5 + \tilde B_{20}(q^2)\,\frac{q^{\{\mu}P^{\nu\}}}{2m_N}\gamma^5\Bigr]u_N(p,s),\,\nonumber\\
    \langle N(p^\prime,s^\prime)| \mathcal{O}_{T}^{\mu\nu\rho}| N(p,s)\rangle =
    \bar u_N(p^\prime,s^\prime)\frac{1}{2}\Bigl[&A_{T20}(q^2)\, i\sigma^{[\mu\{\nu]}P^{\rho\}} + \tilde A_{T20}(q^2)\,\frac{P^{[\mu}q^{\{\nu]}P^{\rho\}}}{m_N^2} + \nonumber\\
   & B_{T20}(q^2)\, \frac{\gamma^{[\mu} q^{\{\nu]}P^{\rho\}}}{2m_N} + \tilde B_{T20}(q^2)\,\frac{\gamma^{[\mu} P^{\{\nu]}q^{\rho\}}}{m_N}\Bigr]u_N(p,s),
  \end{align}
\end{widetext}
where $u_N$ are nucleon spinors. $q=p'-p$ is the momentum transfer,
$P=(p'+p)/2$, and $m_N$ is the nucleon mass.  For zero momentum transfer,
i.e. $p=p'$, we have:

\begin{align}
  \langle N(p, s') |\mathcal{O}_V^{\mu\nu}|N(p,s)\rangle &= \frac{1}{2}A_{20}(0)\llangle\gamma^{\{\mu}p^{\nu\}}\rrangle,\nonumber\\
  \langle N(p, s') |\mathcal{O}_A^{\mu\nu}|N(p,s)\rangle &= \frac{i}{2}\tilde{A}_{20}(0)\llangle\gamma^{\{\mu}p^{\nu\}}\gamma^5\rrangle,\,\nonumber\\
  \langle N(p, s') |\mathcal{O}_T^{\mu\nu\rho}|N(p,s)\rangle &= \frac{i}{2}A_{T20}(0)\llangle\sigma^{[\mu\{\nu]}p^{\rho\}}\rrangle, 
\end{align}
where we use the shorthand notation $\llangle\cdot \rrangle$ to denote
an enclosed quantity between nucleon spinors $\bar{u}_N$ and $u_N$. The generalized
form factors in the forward limit are related to the isovector
momentum fraction, helicity and  transversity moments via
$\langle x \rangle_{u-d}=A_{20}(0)$, $\langle x \rangle_{\Delta u -
  \Delta d} = \tilde{A}_{20}(0)$, and $\langle x \rangle_{\delta u -
  \delta d} = A_{T20}(0)$.

\section{Methodology}
\label{sec:methodology}

\subsection{Gauge ensembles}

We use three gauge ensembles with the parameters listed in
Table~\ref{table:sim}. Two ensembles are generated with two mass
degenerate (\Nfa) up and down quarks with their mass tuned to
reproduce the physical pion mass~\cite{Abdel-Rehim:2015pwa} using two
lattice volumes of $48^3\times96$ and $64^3\times 128$ allowing one to
test for finite volume dependence. We will refer to these ensembles as
the small \Nfa{} ensemble and large \Nfa{} ensemble in addition to the
identifier in the first column of Table~\ref{table:sim}. The third
ensemble is generated on a lattice of $64^3\times
128$~\cite{Alexandrou:2018egz} with two degenerate light quarks and
the strange and charm quarks in the sea (\Nfb) with masses tuned to
reproduce respectively, the physical mass of the pion, kaon and
D$_s$-meson, keeping the ratio of charm to strange quark mass
$m_c/m_s\simeq 11.8$~\cite{Aoki:2019cca}.  For the valence strange and
charm quarks we use Osterwalder-Seiler
fermions~\cite{Osterwalder:1977pc} with masses tuned to reproduce the
mass of the $\Omega^-$ and the $\Lambda^+_c$
baryons~\cite{Alexandrou:2017xwd}, respectively. We will refer to this
ensemble simply as the \Nfb{} ensemble, and to all three ensembles
used as physical point ensembles. The lattice spacing $a$ is
determined using the nucleon mass. The procedure employed to determine
$a$ is outlined in Ref.~\cite{Alexandrou:2018sjm}.

These ensembles use the twisted mass fermion discretization
scheme~\cite{Frezzotti:2000nk,Frezzotti:2003ni} and include a
clover term~\cite{Sheikholeslami:1985ij}.  Twisted mass fermions (TMF)
provide an attractive formulation for lattice QCD allowing for
automatic ${\cal O}(a)$ improvement~\cite{Frezzotti:2003ni} of
physical observables, an important property for evaluating the
quantities considered here. The clover term added to the TMF action
allows for reduced $\mathcal{O}(a^2)$ breaking effects between the
neutral and charged pions~\cite{Abdel-Rehim:2015pwa}. This leads to
the stabilization of physical point simulations while retaining at the
same time the particularly significant ${\cal O}(a)$ improvement that
the TMF action features. For more details on the TMF formulation see
Refs.~\cite{Frezzotti:2004wz,Frezzotti:2005gi,Boucaud:2008xu} and for the simulation
strategy
Refs.~\cite{Chiarappa:2006ae,Abdel-Rehim:2015pwa,Abdel-Rehim:2015owa,Alexandrou:2018egz}.

\begin{widetext}
  \begin{center}   
    \begin{table}[ht!]
      \caption{Simulation parameters for the
        \Nfb~\cite{Alexandrou:2018egz} and
        \Nfa~\cite{Abdel-Rehim:2015pwa} ensembles used in this
        work. When two errors are given, the first error is
        statistical and the second is systematic. The lattice spacing
        is determined using the nucleon mass, as explained in
        Ref.~\cite{Alexandrou:2017xwd} for the cA2.09.48 ensemble and
        in Ref.~\cite{Alexandrou:2018sjm} for the cB211.072.64
        ensemble. For the \Nfa{} ensembles, the systematic error in
        the lattice spacing is due to the fact that the pion mass is
        underestimated and an interpolation is carried out using
        one-loop chiral perturbation theory to interpolate to the
        physical pion mass. More details can be found in
        Ref.~\cite{Alexandrou:2018sjm}. The systematic error in the
        pion mass when expressed in physical units is due to the error
        in the lattice spacing. The volume given in the fifth column
        is in lattice units.}
      \label{table:sim}
      \begin{tabular}{l r@{.}l r@{.}l l  r@{$\times$}l c r@{.}l ccc r@{.}l r}
    \hline\hline
    Ensemble &\multicolumn{2}{c}{$c_{\rm SW}$} & \multicolumn{2}{c}{$\beta$} & \multicolumn{1}{c}{N\textsubscript{f}} & \multicolumn{2}{c}{Vol.} & $m_\pi L$ & \multicolumn{2}{c}{$a$ [fm]} & $m_N/m_\pi$ & $a m_\pi$ & $a m_N$ & \multicolumn{2}{c}{$m_\pi$ [GeV]} &$L$ [fm] \\
    \hline
    cB211.072.64 & 1&69    & 1&778 & 2+1+1 & $64^3$&$128$ & 3.62  & 0&0801(4)    & 6.74(3)    & 0.05658(6)   & 0.3813(19)      & 0&1393(7)          &5.12(3) \\
    cA2.09.64 & 1&57551 & 2&1   & 2        & $64^3$&$128$ & 3.97  & 0&0938(3)(1) & 7.14(4)    & 0.06193(7)   & 0.4421(25)      & 0&1303(4)(2)       &6.00(2) \\
    cA2.09.48 & 1&57551 & 2&1   & 2        & $48^3$&$96$  & 2.98  & 0&0938(3)(1) & 7.15(2)    & 0.06208(2)   & 0.4436(11)      & 0&1306(4)(2)       &4.50(1) \\
    \hline\hline
  \end{tabular}
 \end{table}
  \end{center}
\end{widetext}

\subsection{Correlation functions}
Extraction of the nucleon matrix elements on the lattice proceeds with
the evaluation of two- and three-point correlation functions. All
expressions that follow are in Euclidean space. The three-point
functions are given by
\begin{align}
  C^{\mu\nu}(\Gamma;\vec{q},\vec{p}\,';t_s,t_{\rm ins},t_0) {=} \sum_{\vec{x}_{\rm ins},\vec{x}_s} e^{i (\vec{x}_{\rm ins} {-} \vec{x}_0)  \cdot \vec{q}}  e^{-i(\vec{x}_s {-} \vec{x}_0)\cdot \vec{p}\,'} {\times} \nonumber \\
  \textrm{Tr} \left[ \Gamma \langle J_N(t_s,\vec{x}_s) \mathcal{O}_{\mathcal H}^{\mu\nu}(t_{\rm ins},\vec{x}_{\rm ins}) \bar{J}_N(t_0,\vec{x}_0) \rangle \right],\label{eq:thrp}
\end{align}
where $q=p'-p$ is the momentum transfer. We give the general
expressions for the matrix elements and corresponding correlation
functions using any operator insertion $\mathcal{O}_{\mathcal
  H}^{\mu\nu}$ with $\mu\nu$ arbitrary with the understanding that for
the tensor operator there is an additional index $\rho$.  For the case
of the moment of the tensor PDF, where we need a third index, we will
explicitly include all indices. The initial coordinates $x_0$ are
referred to as the \textit{source} position, $x_{\rm ins}$ as the
\textit{insertion}, and $x_s$ as the \textit{sink}. $\Gamma$ is a
projector acting on spin indices, and we will use either the
unpolarized $\Gamma_0 {=} \frac{1}{2}(1{+}\gamma_0)$ or the three
polarized $\Gamma_k{=}i \gamma_5 \gamma_k \Gamma_0 $ combinations. For
$J_N$, we use the standard nucleon interpolating operator:
\begin{equation}
  J_N(\vec{x},t)=\epsilon^{abc}u^a(x)[u^{\intercal b}(x)\mathcal{C}\gamma_5d^c(x)]\,,
\end{equation}
where $u$ and $d$ are up- and down-quark spinors and
$\mathcal{C}{=}\gamma_0 \gamma_2$ is the charge conjugation
matrix. Inserting a complete set of states in Eq.~(\ref{eq:thrp}), one
obtains a tower of hadron matrix elements with the quantum numbers of
the nucleon multiplied by overlap terms and time dependent
exponentials. For large enough time separations, the excited state
contributions are suppressed compared to the nucleon ground state and
one can then extract the desired matrix element. Knowledge of
two-point functions is required in order to cancel time dependent
exponentials and overlaps. They are given by
\begin{align}
  C(\Gamma_0,\vec{p};t_s,t_0) {=} 
  \sum_{\vec{x}_s} e^{{-}i (\vec{x}_s{-}\vec{x}_0) \cdot \vec{p}}\times& \nonumber\\
  \textrm{Tr}  \left[ \Gamma_0 
    {\langle}J_N(t_s,\vec{x}_s) \bar{J}_N(t_0,\vec{x}_0) {\rangle}
    \right].&
  \label{eq:twop}
\end{align}

In order to increase the overlap of the interpolating operator $J_N$
with the proton state and thus decrease overlap with excited states we
use Gaussian smeared quark
fields via~\cite{Alexandrou:1992ti,Gusken:1989qx},

\begin{align}
  \psi_{\rm smear}^a(t,{\vec{x}}) &= \sum_{\vec{y}} F^{ab}({\vec{x}},{\vec{y}};U(t))\ \psi^b(t,{\vec{y}})\,,\\  
  F &= (\mathbbm{1} + {\alpha} H)^{n} \,, \nonumber\\  
  H({\vec{x}},{\vec{y}}; U(t)) &= \sum_{i=1}^3[U_i(x) \delta_{x,y-\hat\imath} + U_i^\dagger(x-\hat\imath) \delta_{x,y+\hat\imath}],
\end{align}
with APE smearing~\cite{Albanese:1987ds} applied to the gauge fields
$U_\mu$ entering the Gaussian smearing hopping matrix $H$. For the APE
smearing~\cite{Albanese:1987ds} we use 50 iteration steps and
$\alpha_\textrm{APE}{=}0.5$. The Gaussian smearing parameters are
tuned to yield approximately a root mean square radius for the nucleon
of about 0.5~fm, which has been found to yield early convergence of
the nucleon two-point functions to the nucleon mass. This can be
achieved by a combination of the smearing parameters $\alpha$ and
$n$. We use $\alpha$=0.2 and $n$=125 for the \Nfb{} ensemble, and
$\alpha$=0.2 and 4.0 and $n$=90 and 50 for the large and small \Nfa{}
ensembles respectively.

\subsection{Extraction of matrix element}

In order to cancel time dependent exponentials and unknown overlaps of
the interpolating fields with the physical state one constructs
appropriate ratios of three- to two-point functions.  We consider an
optimized ratio constructed such that the two-point functions entering
in the ratio utilize the shortest possible time separation to keep the
statistical noise minimal as well as benefit from correlations. The
ratio~\cite{Alexandrou:2013joa,Alexandrou:2011db,Alexandrou:2006ru}
used is given by
\begin{align}
R^{\mu\nu}(\Gamma;\vec{p}\,',\vec{p};t_s,t_\textrm{ins}) = \frac{C^{\mu\nu}(\Gamma;\vec{p}\,',\vec{p};t_s,t_\textrm{ins})}{C(\Gamma_0;\vec{p}\,';t_s)} \times \nonumber \\
\sqrt{\frac{C(\Gamma_0,\vec{p};t_s{-}t_\textrm{ins}) C(\Gamma_0,\vec{p}\,';t_\textrm{ins}) C(\Gamma_0,\vec{p}\,';t_s)}{C(\Gamma_0,\vec{p}\,';t_s{-}t_\textrm{ins}) C(\Gamma_0,\vec{p};t_\textrm{ins}) C(\Gamma_0,\vec{p};t_s)}} \,,
\label{eq:ratio}
\end{align}
where from now on $t_s$ and $t_\textrm{ins}$ are taken to be relative
to the source $t_0$, i.e. we assume $t_0$=0 without loss of
generality. In the limit of large time separations, $(t_s{-}t_{\rm
  ins}) \gg a$ and $t_\textrm{ins} \gg a$, the lowest state dominates
and the ratio becomes time independent
\begin{equation}
R^{\mu\nu}(\Gamma;\vec{p}\,',\vec{p};t_s,t_{\rm
  ins})\xrightarrow[t_\textrm{ins}\gg a]{t_s-t_\textrm{ins}\gg
  a}\Pi^{\mu\nu}(\Gamma;\vec{p}\,',\vec{p})\,.\label{eq:asymptotic time}
\end{equation}

The generalized form factors are then extracted from linear
combinations of $\Pi^{\mu\nu}(\Gamma;\vec{p}\,',\vec{p})$. In our
approach, we use sequential inversions through the sink, fixing the
sink momentum $\vec{p}^\prime$=0, which implies that the source
momentum is fixed via momentum conservation to
$\vec{p}$=$-\vec{q}$. The general expressions relating
$\Pi^{\mu\nu}(\Gamma;\vec{q})$ to the generalized form factors are
provided in Appendix~\ref{sec:appendix gff extraction}. For the
special case of zero momentum transfer $\vec{q}=0$, the expressions of
Appendix~\ref{sec:appendix gff extraction} simplify to
\begin{align}
  \Pi^{00}_V(\Gamma_0) &= -\frac{3m_N}{4}\langle x \rangle_{u-d},\nonumber\\
  \Pi^{kk}_V(\Gamma_0) &= -\frac{m_N}{4}\langle x \rangle_{u-d},\nonumber\\
  \Pi^{j0}_A(\Gamma_k) &= -\frac{im_N}{2}\delta_{jk}\langle x \rangle_{\Delta u-\Delta d},\nonumber\\
  \Pi^{\mu\nu\rho}_T(\Gamma_k) &= i\epsilon_{\mu\nu\rho k}\frac{m_N}{8}(2\delta_{0\rho}-\delta_{0\mu}-\delta_{0\nu})\langle x \rangle_{\delta u-\delta d},
\end{align}
with $j,k{=}1,2,3$ and $\mu,\nu,\rho{=}0,1,2,3$. All expressions are given in Euclidean space.

The GFFs, given in Appendix~\ref{sec:appendix gff extraction},
depend only on the momentum transfer squared ($Q^2=-q^2$), while
$\Pi^{\mu\nu}(\Gamma;\vec{q})$ depends on $\vec{q}$. The system is
therefore overconstrained, and thus for extracting the GFFs, we form
the matrix $\mathcal{G}$ defined by
\begin{equation}
  \Pi^{\mu\nu}(\Gamma; \vec{q}) = \mathcal{G}^{\mu\nu}(\Gamma; \vec{q}) F(Q^2),
  \label{eq:kin}
\end{equation}
where $\mathcal{G}$ is an array of kinematic coefficients given in the
expressions in Appendix~\ref{sec:appendix gff extraction} and $F$ is
the vector of GFFs. For example for the vector operator ${\cal O}_V^{\mu\nu}$,
$F^\intercal=(A_{20}, B_{20},C_{20})$ and thus $\mathcal{G}$ is an
$N\times 3$ matrix where $N$ is the number of elements contributing to
a given value of $Q^2$. To obtain $F$,
we will use the singular value decomposition (SVD) of $\mathcal{G}$,
combined with three methods for the identification of excited states.

\subsection{Treatment of excited states}
\label{sec:excited states}
Ensuring that the asymptotic behavior of Eq.~(\ref{eq:asymptotic
  time}) holds is a delicate process. This is because the statistical
noise exponentially increases with increasing sink-source separation
$t_s$. In our analysis, we use three methods to study the dependence
of the three- and two-point correlation functions on $t_s$ and
$t_s-t_\textrm{ins}$. This allows us to study the effect of excited
states and thus better identify the convergence to the desired nucleon matrix
element. The methods employed are as follows:

\vspace{0.15cm}
\noindent \emph{Plateau method:} In this method we use the ratio in
Eq.~(\ref{eq:ratio}) in search of a time-independent window (plateau)
and extract a value by fitting to a constant. We then seek convergence of the
extracted plateau value as we increase $t_s$ that then produces the
desired matrix element.

\vspace{0.15cm}
\noindent \emph{Two-state method:} Within this method, we fit the two-
and three-point functions keeping terms up to the first excited state,
namely we use
\begin{equation}
C(\vec{q},t_s) = c_0(\vec{q}) e^{-E_N(\vec{q}) t_s} + c_1(\vec{q}) e^{-E^*(\vec{q}) t_s}\,,
\label{eq:twop twost}
\end{equation}
\begin{align}
  C^{\mu\nu}(\Gamma;\vec{q};t_s,t_{\rm ins}) &= 
   A_{00}^{\mu\nu}(\Gamma,\vec{q}) e^{-m_N(t_s-t_{\rm ins})-E_N(\vec{q})t_{\rm ins}} \nonumber \\
  & + A_{01}^{\mu\nu}(\Gamma,\vec{q}) e^{-m_N(t_s-t_{\rm ins})-E^*(\vec{q})t_{\rm ins}} \nonumber \\
  & + A_{10}^{\mu\nu}(\Gamma,\vec{q}) e^{-m_N^*(t_s-t_{\rm ins})-E_N(\vec{q})t_{\rm ins}} \nonumber \\
  & + A_{11}^{\mu\nu}(\Gamma,\vec{q}) e^{-m_N^*(t_s-t_{\rm ins})-E^*(\vec{q})t_{\rm ins}},
\label{eq:thrp twost}
\end{align}
where $m_N$ ($m_N^*$) and $E_N(\vec{q})$ ($E^*(\vec{q})$) are the mass
and energy of the ground (first excited) state with momentum
$\vec{q}$, respectively. The ground state corresponds to a single
particle, so its energy at finite momentum is given by the continuum
dispersion relation, $E_N(\vec{q})=\sqrt{\vec{q}^2+m_N^2}$, where
$\vec{q}=\frac{2\pi}{L}\vec{n}$ with $\vec{n}$ a lattice vector with
components $n_i\in (-\frac{L}{2a}, \frac{L}{2a}]$. In
  Appendix~\ref{sec:appendix dispersion} we check that the continuum
  dispersion relation is satisfied for all $Q^2$ values considered in
  this work. The first excited state, on the other hand, is allowed to
  be a two-particle state, although we expect the overlap to be volume
  suppressed.  We fit the two-point function at zero momentum and the
  two-point function with momentum $\vec{q}$ yielding the fit
  parameters $m_N$, $m_N^*$, $E^*(\vec{q})$, $c_0(\vec{0})$,
  $c_1(\vec{0})$, $c_0(\vec{q})$, and $c_1({\vec{q}})$.  The
  three-point function is then fitted for the four fit parameters
  $A_{00}$, $A_{01}$, $A_{10}$, and $A_{11}$. For extracting the
  moments in the case of zero momentum transfer, the two-point
  function fit reduces to four parameters and the three-point function
  fit to three. The errors of the fit parameters of the two-point
  functions are propagated by carrying out the fits within the
  resampling method used for each ensemble, i.e. within jackknife for
  the case of the \Nfb{} ensemble for which all results are obtained
  on the same configurations, and within a bootstrap for the \Nfa{}
  ensembles (see Table~\ref{table:stats}). The desired matrix element
  is then given by
\begin{equation}
  \Pi^{\mu\nu}(\Gamma;\vec{q})=\frac{A_{00}^{\mu\nu}(\Gamma,\vec{q})}{\sqrt{c_0(0) c_0(\vec{q})}}.
\label{eq:twost}
\end{equation}

\vspace{0.15cm}
\noindent \emph{Summation method:} Summing over $t_{\rm ins}$ in the
ratio of Eq.~(\ref{eq:ratio}) yields a geometric sum~\cite{Maiani:1987by,Capitani:2012gj} from which we obtain,
\begin{align}
  S^{\mu\nu}(\Gamma;\vec{q};t_s) &= \sum_{t_{\rm ins}=2a}^{t_s-2a} R^{\mu\nu}(\Gamma;\vec{q};t_s,t_{\rm ins}) = \nonumber \\
  &\hspace*{0.35cm}c + \Pi^{\mu\nu}(\Gamma;\vec{q}) {\times}t_s + \mathcal{O}(e^{-(m_N^*-m_N) t_s})
  \label{eq:summation}
\end{align}
where the ground state contribution, $\Pi^{\mu\nu}(\Gamma;\vec{q})$,
is extracted from the slope of a linear fit with respect to $t_s$. The
advantage of the summation method is that, despite the fact that it
still assumes a single state dominance, the excited states are
suppressed exponentially with respect to $t_s$ instead of $t_s-t_{\rm
  ins}$ that enters in the plateau method. On the other hand, the
errors tend to be larger.

For all three methods, we carry out correlated fits to the data, i.e.  we
compute the covariance matrix $v_{ij}$ between jackknife or bootstrap
samples and minimize
\begin{equation}
  \chi_c^2 = [y_i - f(\vec{b}, \{t_s,t_\textrm{ins}\})] v^{-1}_{ij} [y_j - f(\vec{b}, \{t,t_\textrm{ins}\})],
  \label{eq:chisq fit cova}
\end{equation}
where $y_i$ are the lattice data,
i.e. $R^{\mu\nu}(\Gamma;\vec{q};t_s,t_\textrm{ins})$,
$C^{\mu\nu}(\Gamma;\vec{q};t_s,t_\textrm{ins})$, or
$S^{\mu\nu}(\Gamma;\vec{q};t_s)$ depending on whether we are using the
plateau, two-state, or summation method, respectively;
$f(\vec{b},\{t_s,t_\textrm{ins}\})$ is the fit function, which depends
on the variables $t_\textrm{ins}$ and/or $t_s$ according to which of
the three methods we use for the extraction of the matrix element and
$\vec{b}$ is a vector of the parameters being fitted for.

In the most straightforward approach, one minimizes $\chi^2_c$ of
Eq.~(\ref{eq:chisq fit cova}) once for each combination of current
indices $\mu$ and $\nu$, the momentum vectors $\vec{q}$ that
contribute to the same $Q^2$, and the projection matrix $\Gamma$, to
populate the elements of $\Pi^{\mu\nu}(\Gamma;\vec{q})$. Then, a
second minimization is performed to minimize
\begin{equation}
  \chi^2 = \sum_{\mu,\nu,\vec{q} \; \in
    Q^2}\left[\frac{\mathcal{G}^{\mu\nu}(\Gamma, \vec{q})F(Q^2) -
      \Pi^{\mu\nu}(\Gamma; \vec{q})}{w^{\mu\nu}(\Gamma,
      \vec{q})}\right]^2,
\label{eq:chisq fit}
\end{equation}
where $w$ is the statistical error of $\Pi$. Alternatively, the
correlated generalization of Eq.~(\ref{eq:chisq fit}) can be used, in
which the covariance between bootstrap or jackknife samples of
$\Pi^{\mu\nu}(\Gamma;\vec{q})$ are used. Minimizing $\chi^2$ in
Eq.~(\ref{eq:chisq fit}) is equivalent to taking
\begin{equation}
  F = V^\dagger \Sigma^{-1} U^\dagger \tilde{\Pi}
\end{equation}
where
\begin{align}
  \tilde{\Pi}^{\mu\nu}(\Gamma, \vec{q})&\equiv [w^{\mu\nu}(\Gamma, \vec{q})]^{-1}\Pi^{\mu\nu}(\Gamma, \vec{q}),\nonumber\\
  \tilde{\mathcal{G}}^{\mu\nu}(\Gamma, \vec{q})&\equiv [w^{\mu\nu}(\Gamma, \vec{q})]^{-1}\mathcal{G}^{\mu\nu}(\Gamma, \vec{q}),\,\textrm{and}\nonumber\\
  \tilde{\mathcal{G}} &= U\Sigma V.
\end{align}
In the last line, we have used the SVD of $\tilde{\mathcal{G}}$ where
$U$ is a Hermitian $N\times N$ matrix with $N$ the number of
combinations of $\mu$, $\nu$, $\Gamma$ and components of $\vec{q}$
that contribute and $V$ a Hermitian $M \times M$ matrix with $M$ the
number of GFFs, i.e. typically $M\ll N$. $\Sigma$ is the
pseudodiagonal $N \times M$ matrix of the singular values of
$\tilde{\mathcal{G}}$.

As pointed out in Ref.~\cite{Bali:2018zgl}, a more economical approach
arises if one combines the SVD with the fitting procedure. In the case
of correlated fits this is also more robust since it avoids
instabilities. We thus adopt it also here. From Eq.~(\ref{eq:kin}), we
observe that the product $U^\dagger R$, with $R$ the ratio of
Eq.~(\ref{eq:ratio}) (or $C$ in Eq.~(\ref{eq:thrp}) for the case of
the two-state fit method and $S$ in Eq.~(\ref{eq:summation}) for the
case of the summation) is an $N$-length vector of which only the first
$M$ elements contribute to the GFFs. Rather than $N$ fits to the
individual components of
$R^{\mu\nu}(\Gamma;\vec{q};t_s,t_\textrm{ins})$,
$C^{\mu\nu}(\Gamma;\vec{q};t_s,t_\textrm{ins})$, or
$S^{\mu\nu}(\Gamma;\vec{q};t_s)$ we can therefore
perform $M$ fits to the $M$ first elements of the product $U^\dagger
R$, $U^\dagger C$, or $U^\dagger S$. This ``single step'' approach, as
it is referred to in Ref.~\cite{Bali:2018zgl}, will be employed for
the results that follow. We note that the single step approach
produces exactly the same values and errors as analyzing the original
system, i.e. using Eq.~(\ref{eq:chisq fit cova}) to obtain
$\Pi^{\mu\nu}(\vec{q};\Gamma)$ in the first step and then
Eq.~(\ref{eq:chisq fit}) to obtain $F$ in a second step.

\subsection{Evaluation of correlators and statistics}
\label{sec:sim}

For each of the three ensembles, we calculate three- and two-point
functions from multiple randomly chosen source positions. The
three-point functions are calculated for multiple sink-source
separations to study the contribution of excited states. The
statistics are listed in Table~\ref{table:stats}.
 
\begin{table}[ht!]
  \caption{Statistics used for evaluating the three- and two-point
    functions for the three ensembles.  Columns from left to right are
    the sink-source time separation, the number of configurations
    analyzed, the number of source positions per configuration chosen
    randomly and the total number of measurements for each time
    separation. Rows with ``All'' in the first column refer to
    statistics of the two-point function, while the rest indicate
    statistics for three-point functions. For the entries indicated
    with an asterisk ($^*$), three-point functions are only available
    with projector $\Gamma_0$.}
  \label{table:stats}
  \vspace{0.2cm}
  \begin{tabular}{crrr}
    \hline\hline
    $t_s/a$ & \multicolumn{1}{c}{$N_{\rm conf}$} & \multicolumn{1}{c}{$N_{\rm srcs}$} & \multicolumn{1}{c}{$N_{\rm meas}$} \\
    \hline    
    \multicolumn{4}{c}{cB211.072.64: \Nfb, $64^3{\times}128$} \\
    \hline
    \multicolumn{4}{c}{Three-point correlators}\\
     8 & 750 & 1 &  750 \\
    10 & 750 & 2 & 1500 \\
    12 & 750 & 4 & 3000 \\
    14 & 750 & 6 & 4500 \\
    16 & 750 & 16 & 12000 \\
    18 & 750 & 48 & 36000 \\
    20 & 750 & 64 & 48000 \\
    \hline
    \multicolumn{4}{c}{two-point correlators}\\
    All & 750 & 264 & 198000\\
    \hline\hline
    \multicolumn{4}{c}{cA2.09.64: \Nfa, $64^3{\times}128$} \\
    \hline
    \multicolumn{4}{c}{Three-point correlators}\\
    12 & 333 & 16 & 5328 \\
    14 & 515 & 16 & 8240 \\
    16 & 515 & 32 & 16480 \\
    \hline
    \multicolumn{4}{c}{Two-point correlators}\\
    All & 515 & 32 & 16480 \\
    \hline\hline
    \multicolumn{4}{c}{cA2.09.48: \Nfa, $48^3{\times}96$} \\
    \hline
    \multicolumn{4}{c}{Three-point correlators}\\
    10,12,14 & 578 & 16 & 9248 \\
    16$^{*}$       & 530 & 88 & 46640 \\
    18$^{*}$       & 725 & 88 & 63800 \\
    \hline
    \multicolumn{4}{c}{Two-point correlators}\\
    All & 2153 & 100 & 215300\\
    \hline\hline
  \end{tabular}
\end{table}

Since we use sequential inversions through the sink, an additional
inversion is required for each sink-source time separation $t_s$, and
projector. As mentioned, the sink momentum $\vec{p}^\prime$ is set to
zero.  We invert for all four projectors $\Gamma_\mu$, $\mu=0,1,2,3$,
unless otherwise indicated in Table~\ref{table:stats}.  For the \Nfb{}
and small \Nfa{} ensemble, increased statistics are available for
two-point functions compared to three-point functions. This is because
for these two ensembles we have also evaluated disconnected
contributions, which require higher statistics. We use the full set of
available two-point functions here to improve the accuracy of our
two-state fits. The results for the disconnected contributions will
appear in an upcoming publication.

For the efficient inversion of the twisted mass Dirac operator, we use
an appropriately tuned multigrid
algorithm~\cite{Bacchio:2017pcp,Bacchio:2016bwn,Alexandrou:2016izb}.
This is essential for reaching the $\mathcal{O}(10^6)$ inversions per
ensemble listed in Table~\ref{table:stats}.

\begin{figure}
  \includegraphics[width=\linewidth]{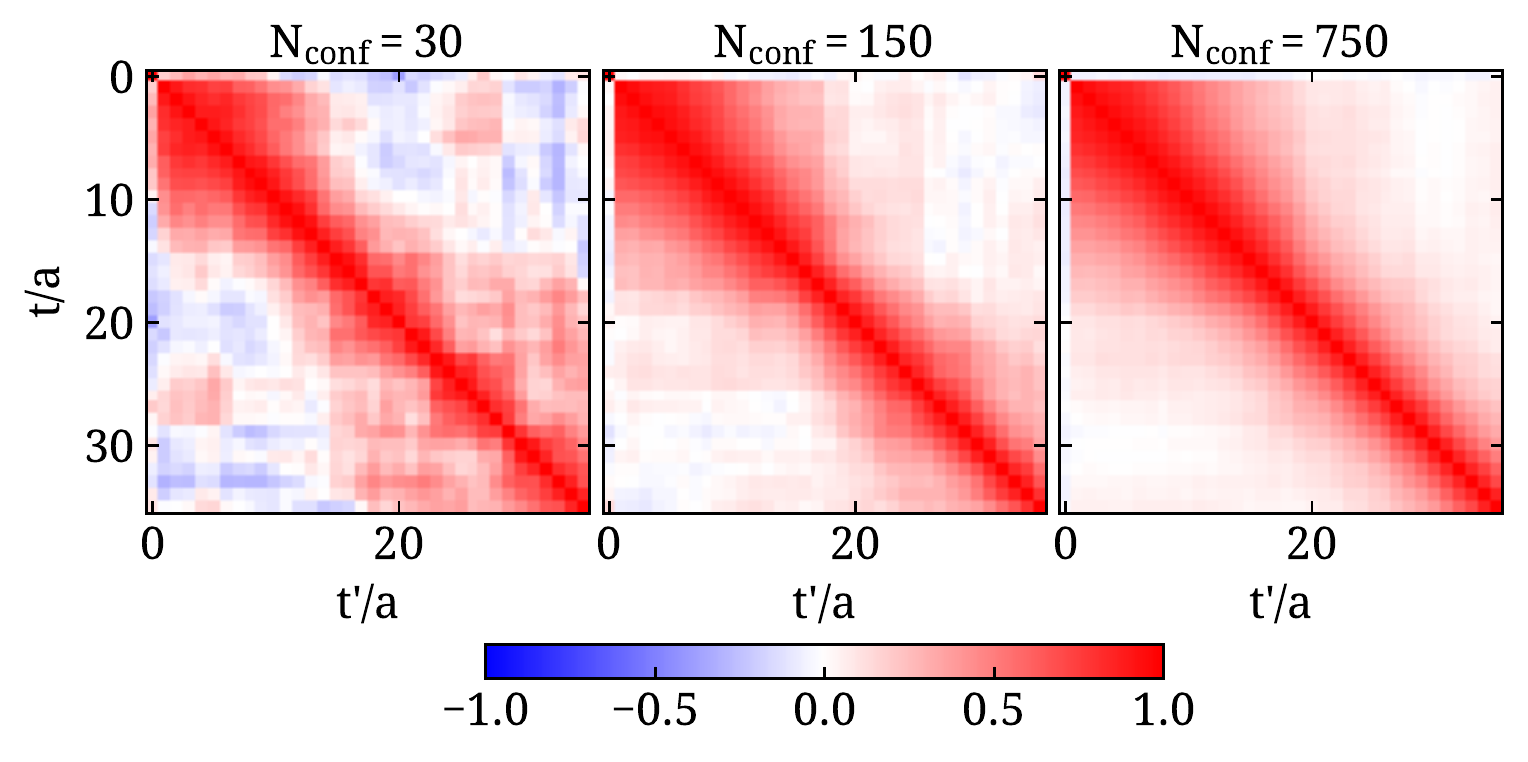}
  \caption{Correlation matrix $\bar{v}_{tt'}$ as defined in
    Eq.~(\ref{eq:cova}) for the case of the two-point correlation
    function for the \Nfb{} ensemble, cB211.072.64, for the first 35
    time slices, i.e.  $t,\,t' \in [0, 35]$. From left to right, we
    show $\bar{v}_{tt'}$ using $N_\textrm{conf}{=}30$, 150, and 750
    configurations.}
  \label{fig:cova}
\end{figure}

It is worth noting that the use of $\chi^2_c$ as defined in
Eq.~(\ref{eq:chisq fit cova}), which takes into account the covariance
of our data in the fit, requires a relatively well conditioned
covariance matrix, which in turn requires high statistics, such as
those listed in Table~\ref{table:stats}. Indicatively, in
Fig.~\ref{fig:cova} we show for the \Nfb{} ensemble the correlation
matrix of the two-point correlation function, defined as
\begin{equation}
  \bar{v}_{tt'} = \frac{v_{tt'}}{\sigma_t \sigma_{t'}} = \frac{\langle [\langle C(t) \rangle - C(t)] [\langle C(t') \rangle - C(t')] \rangle}{\sqrt{\langle C^2(t) \rangle - \langle C(t)\rangle^2} \sqrt{\langle C^2(t') \rangle - \langle C(t')\rangle^2}}
  \label{eq:cova}
\end{equation}
where $\sigma_t$ is the standard deviation of $C(t)$ and all
expectation values are to be taken over
configurations. Fig.~\ref{fig:cova} shows $\bar{v}_{tt'}$ in the range
of time slices used in our analysis, starting from 30 configurations
and quintupling twice to reach the maximum of 750 configurations. As
can be seen, for $N_\textrm{conf}$=750, we obtain a well-defined
covariance, with dominant diagonal and suppressed off-diagonal
fluctuations, as compared to $N_\textrm{conf}$=30 and
$N_\textrm{conf}$=150.

\section{Renormalization functions}
\label{sec:renormalization}

The bare matrix elements of the operators defined in
Eqs.~(\ref{eq:operators}) must be renormalized in order to obtain
physical quantities.  The renormalization functions ($Z$-factors) for
the isovector operators considered here are multiplicative and are
computed nonperturbatively.  We obtain the $Z$-factors using five
ensembles at different values of the pion mass, so that the chiral
limit can be taken. For a proper chiral extrapolation we compute the
$Z$-factors for degenerate quark flavors. For the \Nfa{} we use the
already generated gauge configurations while for the \Nfb{} ensemble
one needs to generate \Nfc{} ensembles with the same $\beta$ value. In
Table~\ref{table:Zmpi} we provide details on the \Nfc{} ensembles used
and the obtained $Z$-factors on each ensemble, while the results for
\Nfa{} ensembles are extensively discussed in
Ref.~\cite{Alexandrou:2015sea}.

We present here a summary of the methodology employed and discuss the
results for the renormalization functions.  We employ the Rome-Southampton method
(RI$'$ scheme)~\cite{Martinelli:1994ty} to compute them
nonperturbatively and impose the conditions
\begin{eqnarray}
   Z_q = \frac{1}{12} {\rm Tr} \left[(S^L(p))^{-1}\, S^{{\rm Born}}(p)\right] \Bigr|_{p^2=\mu_0^2}\,,  \label{Zq_cond}\\[2ex]
   Z_q^{-1}\,Z_{\cal O}\,\frac{1}{12} {\rm Tr} \left[\Gamma^L(p)
     \,\Gamma^{{\rm Born}-1}(p)\right] \Bigr|_{p^2=\mu_0^2} &=& 1\, .
\label{renormalization cond}
\end{eqnarray}
The momentum $p$ is set to the RI$'$ renormalization scale, $\mu_0$,
$S^{{\rm Born}}$ ($\Gamma^{{\rm Born}}$) is the tree-level value of
the fermion propagator (operator), and the trace is taken over spin
and color indices.
 The momentum source method
introduced in Ref.~\cite{Gockeler:1998ye} and employed in
Refs.~\cite{Alexandrou:2010me,Alexandrou:2012mt,Alexandrou:2015sea}
for twisted mass fermions is utilized. This method offers high
statistical accuracy using a small number of gauge configurations. In
this work we use ten configurations to achieve a per mil accuracy. To
reduce discretization effects we use democratic momenta; namely we
consider the same spatial components
\begin{equation}
  (a\,p) \equiv 2\pi \left(\frac{2n_t+1}{2T/a},
\frac{n_x}{L/a},\frac{n_x}{L/a},\frac{n_x}{L/a}\right),\, n_t\in[2, 10],\,n_x\in[2, 5],
\end{equation}
where $T/a$ ($L/a$) is the temporal (spatial) extent of the lattice in lattice units,
and we restrict the momenta up to $(a\,p)^2{\sim} 7$. An important
constraint for the chosen momenta is to suppress the non-Lorentz
invariant contributions $ {\sum_i p_i^4}/{(\sum_i p_i^2
  )^2}{<}0.3$~\cite{Constantinou:2010gr}. This is based on empirical
arguments, as the aforementioned ratio appears in ${\cal O}(a^2)$
terms in the perturbative expressions for the Green's functions, and is
expected to have a non-negligible contribution to higher orders in
perturbation theory (see 
Refs.~\cite{Alexandrou:2010me,Alexandrou:2012mt,Alexandrou:2015sea}
for technical details).
It is worth mentioning that we improve the
nonperturbative estimates by subtracting finite lattice
effects~\cite{Constantinou:2014fka,Alexandrou:2015sea}. The latter are
computed to one-loop in perturbation theory and to all orders in the
lattice spacing, ${\cal O}(g^2\,a^\infty)$. These artifacts are
present in the nonperturbative vertex functions of the fermion
propagator and fermion operators under study.

\begin{center}
    \begin{figure}[!h]
      \centerline{\includegraphics[width=\linewidth]{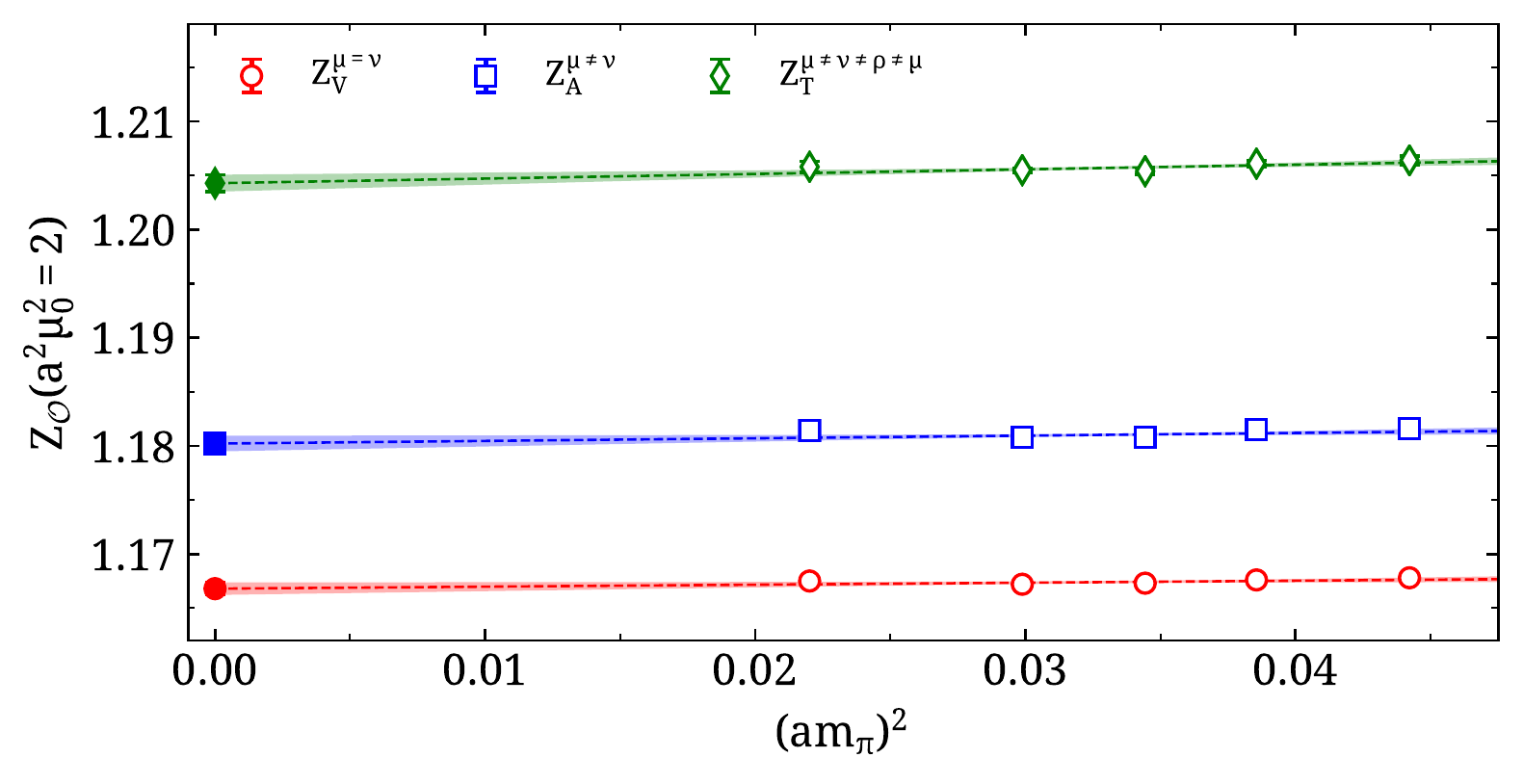}}
      \caption{Example chiral extrapolation of the renormalization
        functions $Z_V^{\mu=\nu}$ (red open circles),
        $Z_A^{\mu\neq\nu}$ (blue open squares), and
        $Z_T^{\mu\neq\nu\neq\rho\neq\mu}$ (green open diamonds) for
        the case of $(a\mu_0)^2=2$. We show with corresponding filled
        symbols the renormalization functions extrapolated to the
        chiral limit, obtained via a quadratic fit, as explained in
        the text.}
      \label{fig:Zchiral}
    \end{figure}
  \end{center}

To obtain the renormalization functions in the chiral limit we perform
an extrapolation using a quadratic fit with respect to the pion mass
of the ensemble, that is,
$a^{\rm{RI}'}\hspace{-0.1cm}(\mu_0)+b^{\rm{RI}'}\hspace{-0.1cm}(\mu_0) \hspace{-0.02cm}
\cdot \hspace{-0.02cm} m_\pi^2$\,, where $a^{\rm{RI}'}$ and
$b^{\rm{RI}'}$ depend on the scheme and the scale. As demonstrated in
our earlier work on the renormalization functions, there is a
negligible dependence on the pion mass~\footnote{Note that the
  renormalization function of the pseudoscalar operator suffers from a
  pion pole and the pion mass dependence is
  significant~\cite{Alexandrou:2015sea}}, which is confirmed by the
results on the \Nfc\ ensembles of Table~\ref{table:Zmpi}. Allowing
$b{\neq}0$ and performing a linear extrapolation with respect to
$m_\pi^2$ the data yield a slope that is compatible with zero within
the small uncertainties. Selected data for all operators are shown in
Table~\ref{table:Zmpi} on each ensemble, at a scale $(a \mu_0)^2{=}2$,
while the chiral extrapolation for this scale is shown in
Fig.~\ref{fig:Zchiral} for the three renormalization functions needed
to renormalize $\langle x \rangle_q$, $\langle x \rangle_{\Delta q}$,
and $\langle x \rangle_{\delta q}$, namely $Z^{\mu=\nu}_{V}$,
$Z^{\mu\neq\nu}_{A}$, and $Z^{\mu\neq\nu\neq\rho\neq\mu}_{T}$
respectively. As can be seen, the pion mass dependence is negligible
and within the statistical uncertainties.

In order to compare lattice values to experimental results one must
convert to the same renormalization scheme and use the same reference
scale $\overline{\mu}$. We employ the commonly used ${\overline{\rm
    MS}}$-scheme at $\overline{\mu}{=}$2 GeV.  The conversion from
RI$'$ to the ${\overline{\rm MS}}$ scheme uses the intermediate
renormalization groupiInvariant (RGI) scheme, which is scale
independent.  Therefore, one may use this property to relate the
renormalization functions between two schemes, and in this case the
RI$'$ and ${\overline{\rm MS}}$:
\begin{align}
Z^{\rm RGI}_{\cal O} =&
Z_{\cal O}^{\mbox{\scriptsize RI$^{\prime}$}} (\mu_0) \, 
\Delta Z_{\cal O}^{\mbox{\scriptsize RI$^{\prime}$}}(\mu_0) \nonumber\\
= &
Z_{\cal O}^{\overline{\rm MS}} (2\,{\rm GeV}) \,
\Delta Z_{\cal O}^{\overline{\rm MS}} (2\,{\rm GeV})\,.
\end{align}

The conversion factor can be extracted from the above relation
\begin{equation}
C_{\cal O}^{{\rm RI}',{\overline{\rm MS}}}(\mu_0,2\,{\rm GeV}) \equiv 
\frac{Z_{\cal O}^{\overline{\rm MS}} (2\,{\rm GeV})}{Z_{\cal O}^{{\rm RI}'} (\mu_0)} = 
\frac{\Delta Z_{\cal O}^{\mbox{\scriptsize RI$^{\prime}$}}(\mu_0)}
     {\Delta Z_{\cal O}^{\overline{\rm MS}}(2\,{\rm  GeV})}\,.
     \label{eq:Conv}
\end{equation}
The quantity $\Delta Z_{\cal O}^{\mathcal S}(\mu_0)$ is expressed in terms of the $\beta$-function and the anomalous dimension $\gamma_{\cal O}^S \equiv \gamma^S$ of the operator
\begin{align}
\Delta Z_{\cal O}^{\mathcal S} (\mu) =&
  \left( 2 \beta_0 \frac {{g^{\mathcal S} (\mu)}^2}{16 \pi^2}\right)
^{-\frac{\gamma_0}{2 \beta_0}}\times\nonumber\\
  &\exp \left \{ \int_0^{g^{\mathcal S} (\mu)} \! \mathrm d g'
  \left( \frac{\gamma^{\mathcal S}(g')}{\beta^{\mathcal S} (g')}
   + \frac{\gamma_0}{\beta_0 \, g'} \right) \right \}\,.
\end{align}

\begin{widetext}
   \begin{center}   
     \begin{table}[h!]
       \caption{Parameters and resulting $Z$-factors for the \Nfc{}
         ensembles needed for the renormalization of the \Nfb{}
         ensemble (cB211.072.64). The first column is the twisted bare
         mass parameter and the second and third columns are the pion
         mass in lattice and physical units respectively. The
         remaining columns give the $Z$-factors in the RI$'$ scheme at
         $(a\mu_0)^2{=}2$. The number in the parentheses is the
         statistical error.}\label{table:Zmpi}
    \begin{tabular}{cccccccccc}
      \hline\hline
      \multicolumn{10}{c}{$\beta$=1.778, $a$=0.0801(4)~fm, ($L^3\times T$) = (24$^3\times$48)} \\\hline
      $a\mu$ & $a m_{\pi}$ & $m_{\pi}$ [MeV] & $Z^{\mu=\nu}_{V}$ & $Z^{\mu\neq\nu}_{V}$ & $Z^{\mu=\nu}_{A}$ & $Z^{\mu\neq\nu}_{A}$ & $Z^{\mu\neq\nu=\rho}_{T}$ & $Z^{\mu\neq\nu\neq\rho\neq\mu}_{T}$ & $Z^{\mu=\nu\neq\rho}_{T}$ \\

      \hline
      0.0060 & 0.14836 & 366(2) & 1.1675(3)        & 1.1835(4)        & 1.1921(4)        & 1.1814(4)        & 1.1863(4)        & 1.2058(5)        & 1.1527(3)        \\
      0.0075 & 0.17287 & 427(2) & 1.1672(2)        & 1.1830(2)        & 1.1917(2)        & 1.1808(2)        & 1.1860(3)        & 1.2055(2)        & 1.1527(2)        \\
      0.0088 & 0.18556 & 458(2) & 1.1673(2)        & 1.1831(2)        & 1.1918(2)        & 1.1808(2)        & 1.1860(2)        & 1.2054(3)        & 1.1528(2)        \\
      0.0100 & 0.19635 & 485(2) & 1.1676(3)        & 1.1836(3)        & 1.1922(3)        & 1.1815(3)        & 1.1866(2)        & 1.2061(3)        & 1.1530(1)        \\
      0.0115 & 0.21028 & 519(3) & 1.1678(3)        & 1.1839(4)        & 1.1927(3)        & 1.1816(4)        & 1.1869(4)        & 1.2064(4)        & 1.1534(3)        \\
      \hline\hline
    \end{tabular}
 \end{table}
 \end{center}
\end{widetext}

The expressions for the one-derivative operators are known to
three-loops in perturbation theory and can be found in
Ref.~\cite{Alexandrou:2015sea} (and references therein).

\begin{center}
    \begin{figure}[!h]
      \centerline{\includegraphics[width=\linewidth]{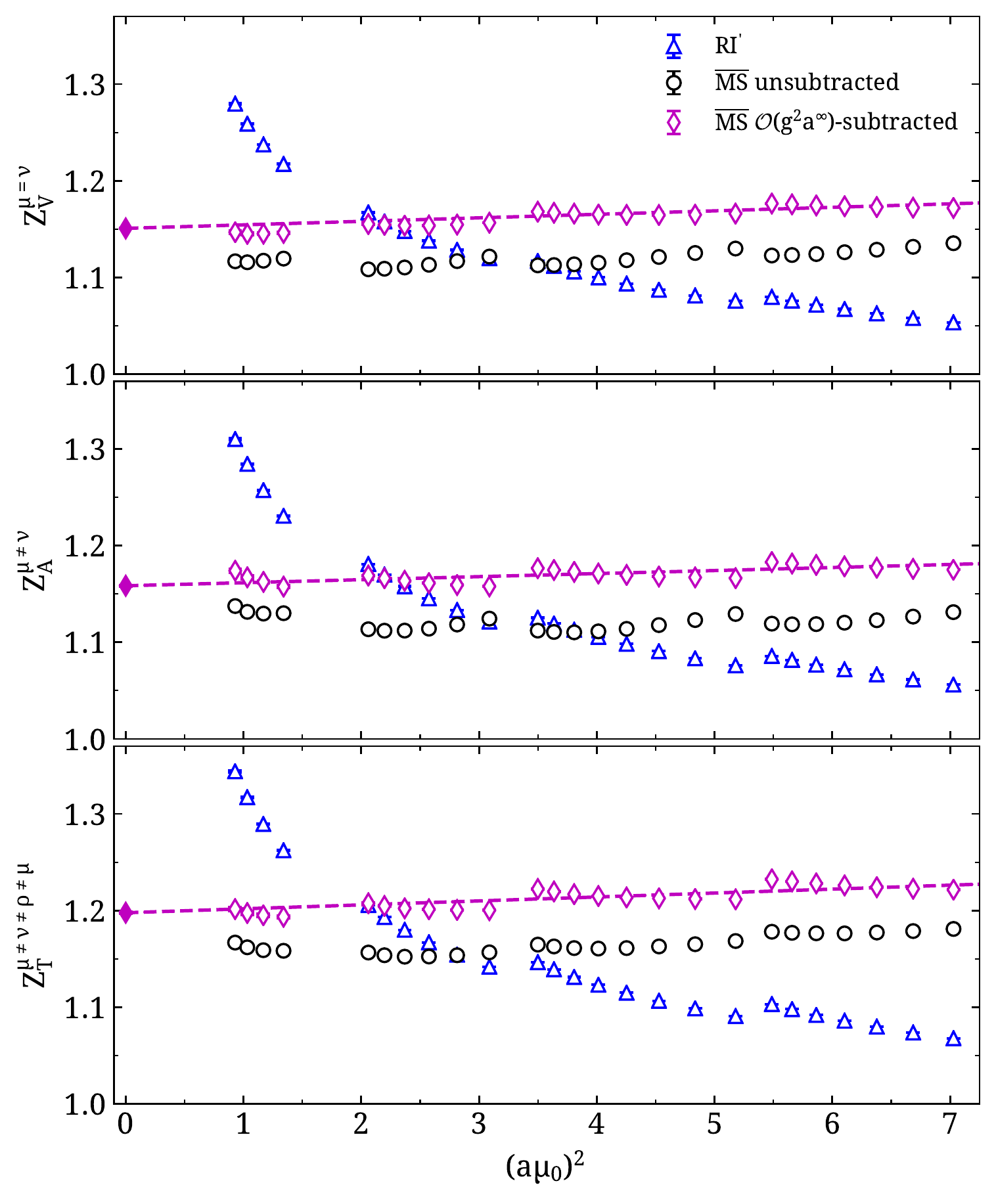}}
      \caption{Chirally extrapolated results for $Z^{\mu=\nu}_{V}$
        (upper plot), $Z^{\mu\neq\nu}_{A}$ (center plot), and
        $Z^{\mu\neq\nu\neq\rho\neq\mu}_{T}$ (lower plot), which are
        needed for $\langle x \rangle_q$, $\langle x \rangle_{\Delta
          q}$, and $\langle x \rangle_{\delta q}$, respectively. The
        data for the RI$'$ scheme are shown with blue triangles, the
        purely nonperturbative data for the ${\overline{\rm MS}}$
        scheme are shown with black circles, and the improved
        ${\overline{\rm MS}}$ estimates with magenta diamonds. The
        data are plotted as a function of the initial renormalization
        scale $(a\,\mu_0)^2$. The dashed lines correspond to the fit
        of Eq.~(\ref{Zfinal}), and the filled magenta diamonds
        represent the final estimate $Z_{\cal O}^{(0)}$.}
      \label{fig:ZRIMS}
    \end{figure}
  \end{center}

\vspace*{0.25cm} In Fig.~\ref{fig:ZRIMS} we compare the
renormalization functions in the RI$'$ and ${\overline{\rm MS}}$
schemes as a function of the RI$'$ renormalization scale,
$\mu_0$. Note that the values in ${\overline{\rm MS}}$ have been
evolved to 2~GeV, and from the plot we can see that the purely
nonperturbative data (black points) exhibit a residual dependence on
$\mu_0$ (the scale they were evolved from, using the appropriate
expressions of Eq.~(\ref{eq:Conv})). This dependence is removed via
two procedures: \\ \text{1.} the subtraction of finite-$a$ effects to
${\cal O}(g^2 a^\infty)$, \\ \text{2.} the extrapolation of $(a\mu_0)^2$
to zero, using the Ansatz
\begin{equation}
Z_{\cal O}(a\,p) = Z_{\cal O}^{(0)} + Z_{\cal O}^{(1)}\cdot(a\,\mu_0)^2\,.
\label{Zfinal}
\end{equation}
$Z_{\cal O}^{(0)}$ corresponds to our final value of the
renormalization functions for operator ${\cal O}$ (filled magenta
diamonds at $(a\,\mu_0)^2{=}0$), and in the above fit we consider
momenta $(a\,\mu_0)^2 {\ge} 2$ for which perturbation theory is
trustworthy and lattice artifacts are still under control.

In Table~\ref{table:renorm} we report our chirally extrapolated values
for the renormalization functions used in this work. The statistical
and systematic uncertainties are given in the first and second
sets of parentheses, respectively. The source of systematic error is related to the
$(a\,\mu_0)^2{\to} 0$ extrapolation and it is obtained by varying the lower and higher fit ranges  between $a\mu_0=2$ and 7 and taking the largest deviation
as the systematic error.
The values given in Table~\ref{table:renorm} are determined using the fit interval $(a\,\mu_0)^2\, \epsilon\, [2-7]$.

\begin{widetext}
  \begin{center}
    \begin{table}[h]
      \caption{Renormalization functions for the operators used in our
        GFF calculation in the $\overline{\textrm{MS}}$ scheme at an energy scale of
        2~GeV. The first row for the \Nfb{} ensemble with
        $\beta$=1.778, and the second row for the two \Nfa{} ensembles
        with $\beta$=2.1.  The number in the first parentheses is the
        statistical error, while the number in the second parentheses
        corresponds to the systematic error obtained by varying the
        fit range in the $(a\,\mu_0)^2{\to} 0$ extrapolation.
        \vspace*{0.25cm}
      }\label{table:renorm}
      \begin{tabular}{cccccccc}
        \hline\hline\\[-2.5ex]
        Ensemble & $Z^{\mu=\nu}_{V}$ & $Z^{\mu\neq\nu}_{V}$ & $Z^{\mu=\nu}_{A}$ & $Z^{\mu\neq\nu}_{A}$ & $Z^{\mu\neq\nu=\rho}_{T}$ & $Z^{\mu\neq\nu\neq\rho\neq\mu}_{T}$ & $Z^{\mu=\nu\neq\rho}_{T}$ \\[0.5ex]
        \hline\\[-2.5ex]
        cB211.072.64     &$\,\,\,$ 1.151(1)(4)   &$\,\,\,$ 1.160(1)(3)     &$\,\,\,$ 1.172(1)(4)     &$\,\,\,$ 1.159(1)(2)  &$\,\,\,$ 1.182(1)(2) &$\,\,\,$ 1.198(1)(5) &$\,\,\,$ 1.154(1)(9) \\
        cA2.09.\{48,64\} &$\,\,\,$ 1.125(3)(2)   &$\,\,\,$  1.140(2)(1)     &$\,\,\,$ 1.149(1)(1)    &$\,\,\,$ 1.136(2)(20) &$\,\,\,$ 1.138(16)(1)   &$\,\,\,$ 1.147(12)(5) &$\,\,\,$          \\[0.5ex]
        \hline\hline
      \end{tabular}
    \end{table}
  \end{center}
\end{widetext}

\section{Results}
\label{sec:results}

\subsection{Zero momentum transfer}
\label{sec:results moments}

\begin{figure}
  \includegraphics[width=\linewidth]{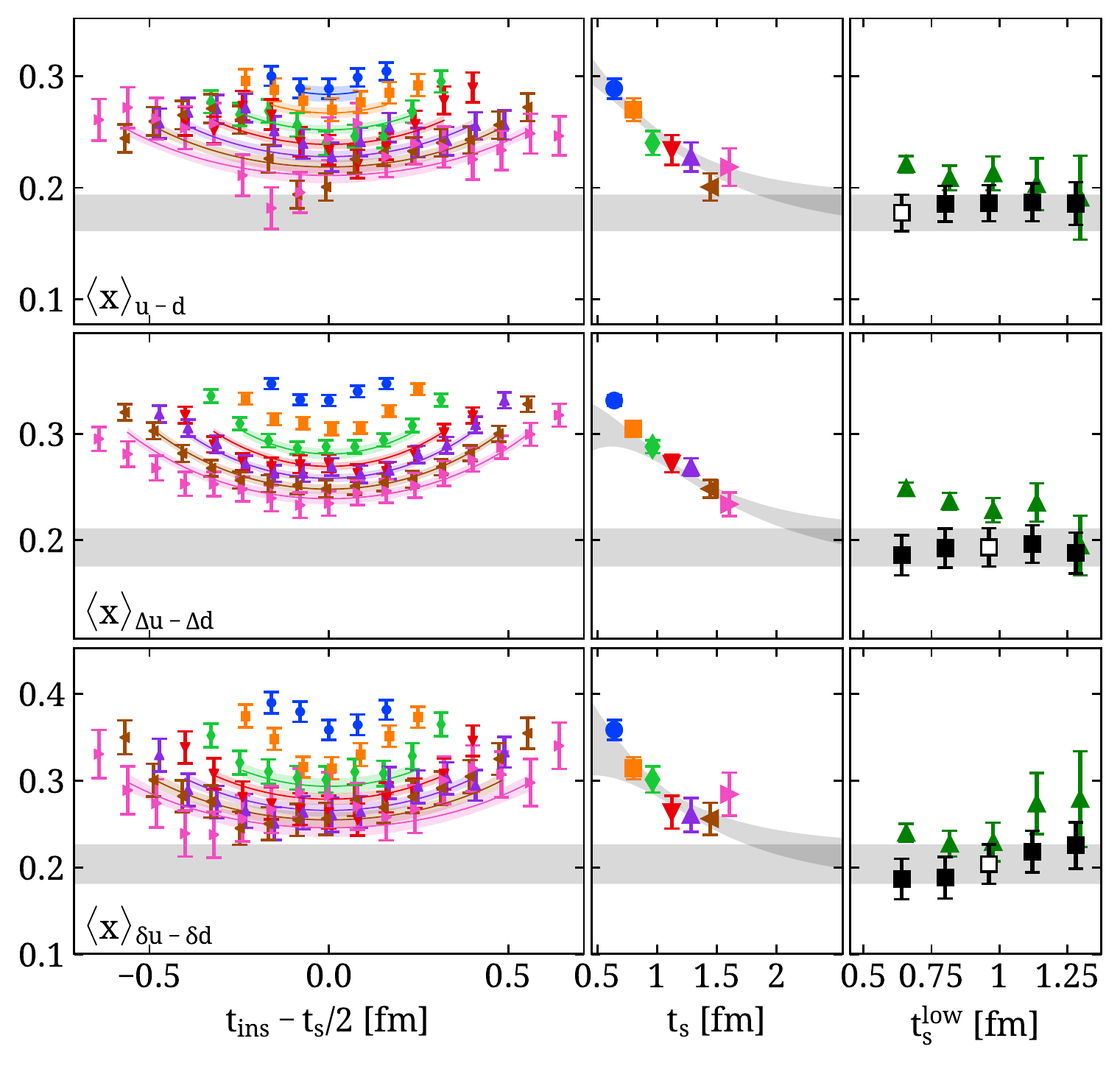}
  \caption{Results for the \Nfb{} ensemble, cB211.072.64, for the
    isovector momentum fraction $\langle x \rangle_{u-d}$ (top row),
    the helicity moment $\langle x \rangle_{\Delta u-\Delta d}$
    (middle row), and the transversity $\langle x \rangle_{\delta
      u-\delta d}$ (bottom row) as a function of $t_s$ or
    $t_s^\textrm{low}$ in physical units. In the left column, we show
    the ratio of Eq.~(\ref{eq:ratio}) for sink-source separation
    $t_s=8a$ (blue circles), $10a$ (orange squares), $12a$ (green
    diamonds), $14a$ (red downwards pointing triangles), $16a$ (purple
    upwards pointing triangles), $18a$ (brown left pointing
    triangles), and $20a$ (magenta right pointing triangles), plotted
    against the insertion time shifted by $t_s/2$ so that the
    midpoints coincide. Data are additionally slightly shifted
    horizontally to ease legibility of overlapping points. The curves
    and corresponding bands are the result of the two-state fit. In
    the middle column, we show the result of the plateau fit for each
    sink-source separation, using the symbol notation of the left
    column. The band is obtained from the two-state fit parameters, as
    explained in the text. The right column shows the result of the
    summation method (green triangles) and the two-state fit method
    (black squares) as a function of the lower value $t_s^{\rm low}$
    included in the fit. The open black square shows the selected
    value and the horizontal band spanning all three columns is the
    associated error band.}
  \label{fig:fits}
\end{figure}

We begin by presenting our results for zero momentum transfer, which
yield the isovector moments of PDFs, i.e. the momentum fraction
$\langle x\rangle_{u-d}$, the helicity $\langle x\rangle_{\Delta
  u-\Delta d}$, and the transversity $\langle x\rangle_{\delta
  u-\delta d}$.  In Fig.~\ref{fig:fits} we show a summary of the
analyses carried out, as described in Sec.~\ref{sec:excited
  states}, for the case of the \Nfb{} ensemble.

In the first column of Fig.~\ref{fig:fits}, we plot the ratios of
Eq.~(\ref{eq:ratio}) for the three moments. In the central column, we
plot the values obtained from plateau fits to the ratio as a function
of the sink-source separation $t_s$.  We show the plateaus obtained
taking the insertion fit range: $t_\textrm{ins}\in[\tau_\textrm{plat},
  t_s-\tau_\textrm{plat}]$ choosing $\tau_\textrm{plat}$ such that
when it is increased, the values obtained for the plateau fit do not
change for each $t_s$. We find $\tau_\textrm{plat}=7a$ satisfies this
criterion, and for the separations for which $t_s<14a$ we plot the
value of the ratio at the midpoint, i.e. for $t_\textrm{ins}=t_s/2$,
in the central column of Fig.~\ref{fig:fits}.

As explained for the two-state fit we first fit the two-point function
at zero momentum. The values we extract for $m_N$ and $m_N^*$ remain
unchanged within errors if we include a second excited state, i.e. if
we perform a three-state fit. While in the spectral decomposition of
the two-point correlation function, the energy state above the nucleon
should include a pion-nucleon with relative momentum, it is noteworthy
that for all ensembles we find a value for $m_N^*$ that is consistent
to the mass of the Roper rather than a multiparticle state, as shown
in Table~\ref{table:m1}.
\begin{table}[h]
  \caption{Values of the excited state mass $m_N^*$, in GeV, as
    extracted from the two-state (first row of results) and
    three-state (second row of results) fits to the two-point
    functions of the three ensembles analyzed in this
    work.}\label{table:m1}
  \begin{tabular}{rr@{.}lr@{.}lr@{.}l}
    \hline\hline
            & \multicolumn{2}{c}{cB211.072.64}  &   \multicolumn{2}{c}{cA2.09.48}     &  \multicolumn{2}{c}{cA2.09.64}    \\\hline
    Two-state:   & 1&43(7)  &  1&59(6)   &  1&45(11) \\
    Three-state: & 1&38(12) &  1&44(15)  &  1&15(16) \\\hline\hline
  \end{tabular}
\end{table}
  
The results obtained using the summation and two-state fit methods are
shown in the right column of Fig.~\ref{fig:fits}, as a function of the
smallest sink-source separation used in the fit
$t_s^\textrm{low}$. For the two-state fit method, we choose the
fit range for the two-point function by requiring the ground state
mass extracted with the two-exponential ansatz of Eq.~(\ref{eq:twop
  twost}) to agree with that obtained from a constant fit to the
effective mass, within half the error of the latter. This analysis
yields $t_s\in [8a,\,35a]$ in the case of the \Nfb{} ensemble and this
is used throughout. Furthermore, we find that taking
$t_\textrm{ins}\in[\tau,\,t_s-\tau]$ for $\tau \ge 3a$ in the
two-state fit yields consistent results, and thus we fix $\tau{=}3a$.

From the right column of Fig.~\ref{fig:fits}, we see that in general
the two-state fit results are stable for all $t_s^\textrm{low}$
values, with the summation method converging as $t_s^\textrm{low}$ is
increased. The bands in the left column of Fig.~\ref{fig:fits} show
the ratio of Eq.~(\ref{eq:ratio}) when using the parameters of the
two-state fit to reproduce the two- and three-point functions, namely
Eqs.~(\ref{eq:twop twost}) and~(\ref{eq:thrp twost}). We see that in
all cases, the predicted bands reproduce the data well.

The band in the central column of Fig.~\ref{fig:fits} is not a fit to
the data; it is drawn using the parameters obtained from the two-state
fit as a function of continuous values for $t_s$ and taking
$t_\textrm{ins}=t_s/2$. The left and central columns show that the
data are reproduced well with the two-state fit ansatz. Furthermore,
the band drawn in the central column reveals that sink-source
separations beyond $\approx 2$~fm are required to obtain plateaus that would
sufficiently suppress the first excited state and therefore yield
agreement between the plateau and two-state fit methods. Such a
separation would not be feasible with currently available
computational resources. Indeed, between our smallest and largest
separations of 0.64~fm and 1.6~fm respectively, we increase statistics
by 64$\times$ (see Table~\ref{table:stats}) while errors increase by
$\sim$2.5$\times$, indicating that to obtain at $\sim 2$~fm the same error
as that obtained at 1.6~fm we would require ${\cal O}(100)$  more
statistics. We will therefore quote the result of the two-state fit
method as our final result, shown by the horizontal band spanning all
columns in Fig.~\ref{fig:fits}.

To choose the $t_s^\textrm{low}$ of the two-state fit for quoting our
final result, we will demand that this agrees with the converged value
of the summation method. For $\langle x\rangle_{u-d}$ the two-state
fit result with $t_s^\textrm{low}{=}8a$ agrees with the result of the
summation method for $t_s^\textrm{low}>$1~fm. We therefore take the
two-state fit result with $t_s^\textrm{low}{=}8a$ as our final value
for the momentum fraction. For the helicity $\langle x\rangle_{\Delta
  u-\Delta d}$ and tensor charge $\langle x\rangle_{\delta u-\delta
  d}$, as can be seen, we need to increase $t_s^{\rm low}$ further to
achieve agreement with the summation method. We therefore take the
value when fitting from $t_s^\textrm{low}{=}12a$ as our final result.

\begin{figure}
  \includegraphics[width=\linewidth]{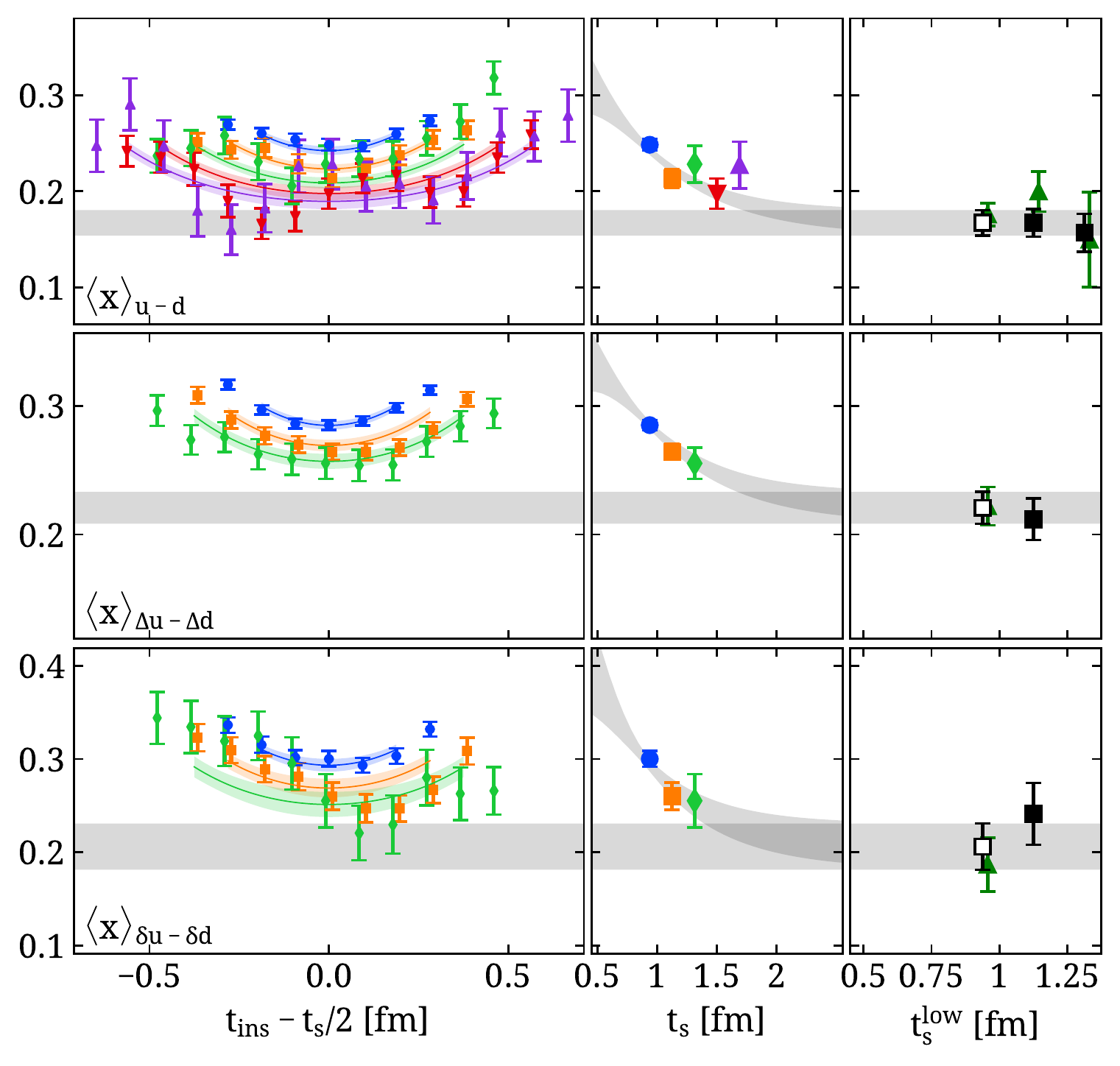}
  \caption{Results for the small \Nfa{} ensemble, cA2.09.48. In the
    left column, we show the ratio of Eq.~(\ref{eq:ratio}) for
    sink-source separation $t_s{=}10a$ (blue circles), $12a$ (orange
    squares), $14a$ (green diamonds), $16a$ (red downwards
    pointing triangles), and $18a$ (purple upwards pointing
    triangles). The rest of the notation is as in
    Fig.~\ref{fig:fits}.}\label{fig:fits cA2.09.48}
\end{figure}

The same analysis is carried out for the small and large \Nfa{}
ensembles, shown in Figs.~\ref{fig:fits cA2.09.48} and~\ref{fig:fits
  cA2.09.64}, respectively. The analysis of the \Nfb{} ensemble, for
which we use seven values of $t_s$ with increased statistics, has
clearly revealed that excited state effects die out slowly and that
one needs to go to larger values of $t_s$~\cite{Bar:2018xyi} keeping
statistical errors small to see clear convergence as also demonstrated
in Ref.~\cite{vonHippel:2016wid}. With this hindsight, we reanalyze
the \Nfa\ ensembles.  For determining the fit ranges of the two-point
function we use the same criteria as for the \Nfb{} ensemble. We find
that $t_s\in [6a,\,24a]$ for the small \Nfa{} ensemble and $t_s\in
[5a,\,24a]$ for the large \Nfa{} ensemble satisfy the agreement
between the values extracted from one-state (plateau) and two-state
fits. While for the \Nfb{} and small \Nfa{} ensembles we have
increased statistics for the two-point functions used in the two-state
fit method, for the large \Nfa{} ensemble we are limited to the same
statistics for two-point functions as those for the three-point
function. The reason is that for the latter ensemble we did not
compute disconnected contributions. This also explains why the lower
fit range for the large \Nfa{} ensemble is smaller as compared to the
small \Nfa{} ensemble, since the two-point correlator has lower
precision.  In the case of the three-point function, for both these
ensembles, in general only three sink-source separations are
available, which allow for only a single point for the summation
method and two points for the two-state fit method. The unpolarized
projector for the case of the small \Nfa{} ensemble is the only
exception, namely for this case we obtain $\langle x \rangle_{u-d}$,
for two additional separations.

\begin{figure}
  \includegraphics[width=\linewidth]{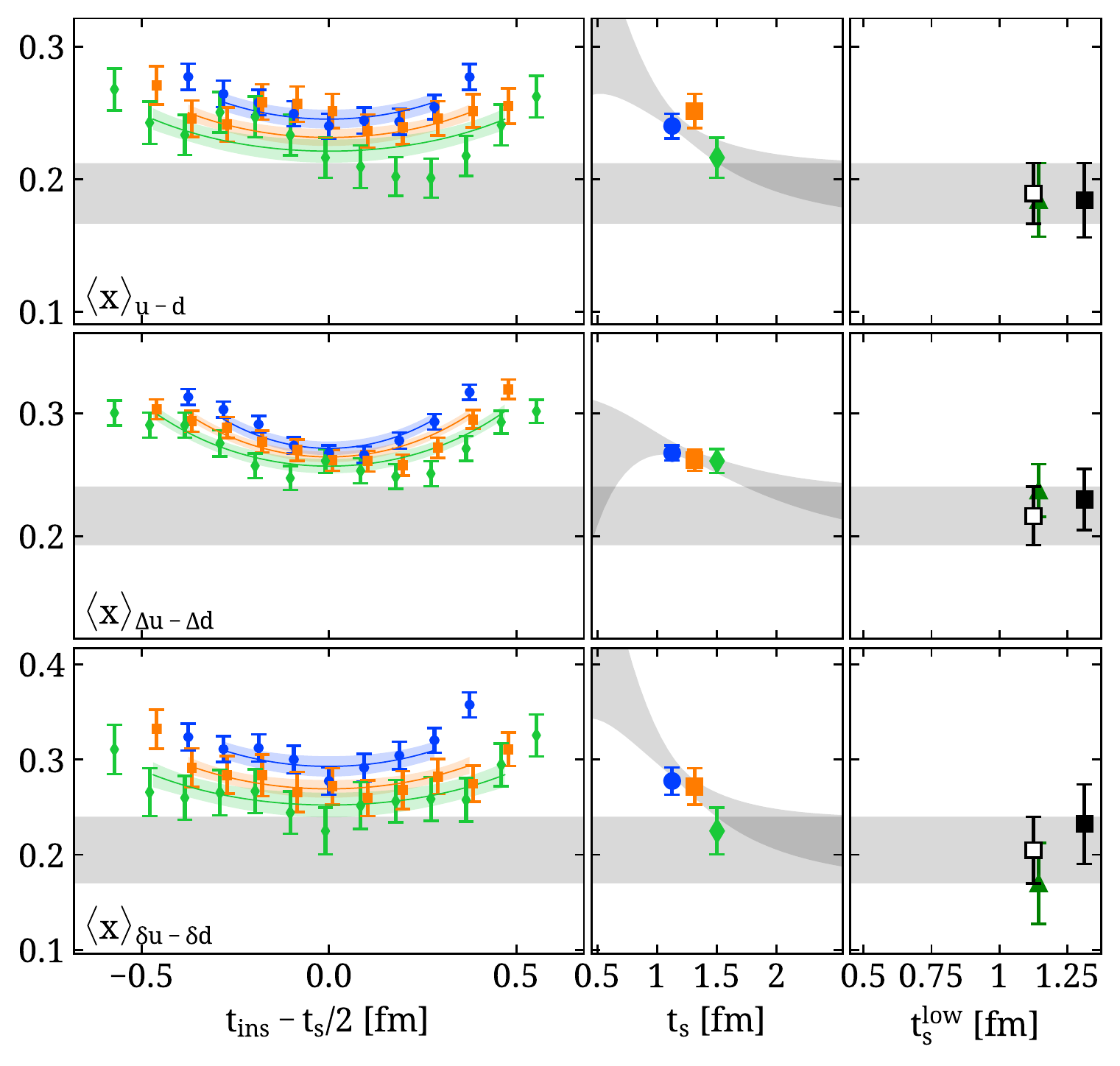}
  \caption{Results for the large \Nfa{} ensemble, cA2.09.64. The left
    column shows the ratio of Eq.~(\ref{eq:ratio}) for sink-source
    separation $t_s{=}12a$ (blue circles), $14a$ (orange squares), and
    $16a$ (green diamonds). The rest of the notation is as in
    Fig.~\ref{fig:fits}.}\label{fig:fits cA2.09.64}
\end{figure}

From Figs.~\ref{fig:fits cA2.09.48} and~\ref{fig:fits cA2.09.64} we
observe a curvature in the ratio data similar to that of the \Nfb{}
ensemble. For the plateau fits shown in the central columns, we use
$\tau_\textrm{plat}=5a$ for both ensembles, determined using the same
criterion as for the \Nfb{} case. Comparing two-state fit and
summation methods, we note that at the smallest $t_s^\textrm{low}$
available for these two ensembles, which is around $\sim$1~fm, we see
agreement between two-state and summation methods. For all three
ensembles, therefore, the summation method at around
$t_s^\textrm{low}\simeq$1~fm converges to the two-state fit result
within errors. We take the two-state fit result as our final value for
these two ensembles. Our final results for the three moments are given
in Table~\ref{table:moments}.

\begin{table}
  \caption{Results for the three isovector moments from the three
    ensembles analyzed in this work. The results are obtained from the
    two-state fit as explained in the text.}\label{table:moments}
  \begin{tabular}{llll}
    \hline\hline
    \multicolumn{1}{c}{Ensemble} & \multicolumn{1}{c}{$\langle x\rangle_{ u- d}$}  & \multicolumn{1}{c}{$\langle x\rangle_{\Delta u- \Delta d}$}  & \multicolumn{1}{c}{$\langle x\rangle_{\delta u-\delta d}$} \\
    \hline
    cB211.072.64  &  0.178(16)  & 0.193(18)  & 0.204(23)  \\
    cA2.09.48     &  0.167(13)  & 0.221(12)  & 0.206(25)  \\
    cA2.09.64     &  0.189(23)  & 0.217(24)  & 0.205(35)  \\
    \hline\hline
  \end{tabular}
\end{table}

Comparing the three moments between the three ensembles, we see in
general that these agree within our statistical errors, an exception
being $\langle x\rangle_{\Delta u- \Delta d}$ for \Nfb{} and the
\Nfa{} ensembles, where agreement is within 1.5$\sigma$ of the former.

Comparing the results obtained using the two \Nfa{} ensembles, which
differ only in their volume, with $m_\pi L$=2.98 to 3.97, respectively
reveals no finite volume effects within our statistical errors for all
three moments. The \Nfb{} ensemble has $m_\pi L$=3.62 (see
Table~\ref{table:sim}), which is between the two volumes with \Nfa{}
and thus we also expect that volume effects are also within the
statistical errors for this ensemble as well. The \Nfb{} ensemble has
a smaller lattice spacing and includes the strange and charm quarks in
the sea. We find that the moments obtained using the \Nfb{} ensemble
and the two \Nfa{} ensembles are in agreement. This suggests that
unquenching effects and cutoff effects for these quantities, at least
within the range of these two lattice spacings, are also smaller than
our statistical uncertainties.
    
\subsection{Finite momentum transfer}
\label{sec:results gffs}

  \begin{figure*}
      \includegraphics[width=\linewidth]{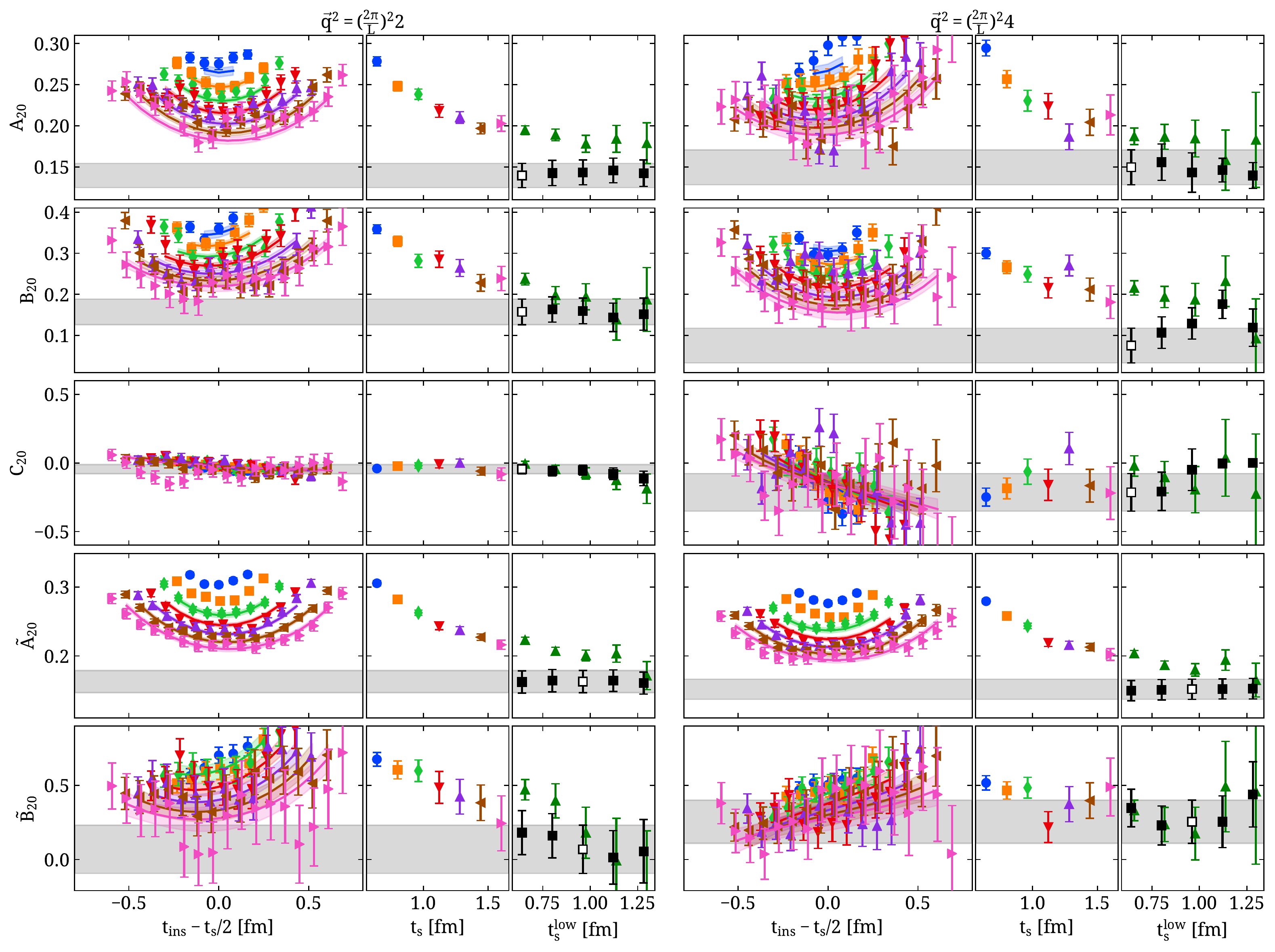}
      \caption{Results for the vector and axial GFFs for the \Nfb{}
        ensemble, cB211.072.64 for two representative $\vec{q}^2$
        values corresponding to $Q^2=0.114$~GeV$^2$ (left) and
        0.222~GeV$^2$ (right). The first three rows show results for
        the three vector GFFs, namely $A_{20}$, $B_{20}$, and
        $C_{20}$, and the last two rows for the two axial GFFs
        $\tilde{A}_{20}$ and $\tilde{B}_{20}$. For each of the two
        values of $Q^2$ shown, we use the same notation as for the
        $Q^2=0$ case of Fig.~\ref{fig:fits}, namely showing the ratio
        obtained as explained in the text (left columns), the result
        of fitting the plateau in the single step approach (central
        columns), and the results from two-state fits and summation
        method (right columns).}\label{fig:fits qsq 2 4}
  \end{figure*}

  \begin{figure*}
      \includegraphics[width=\linewidth]{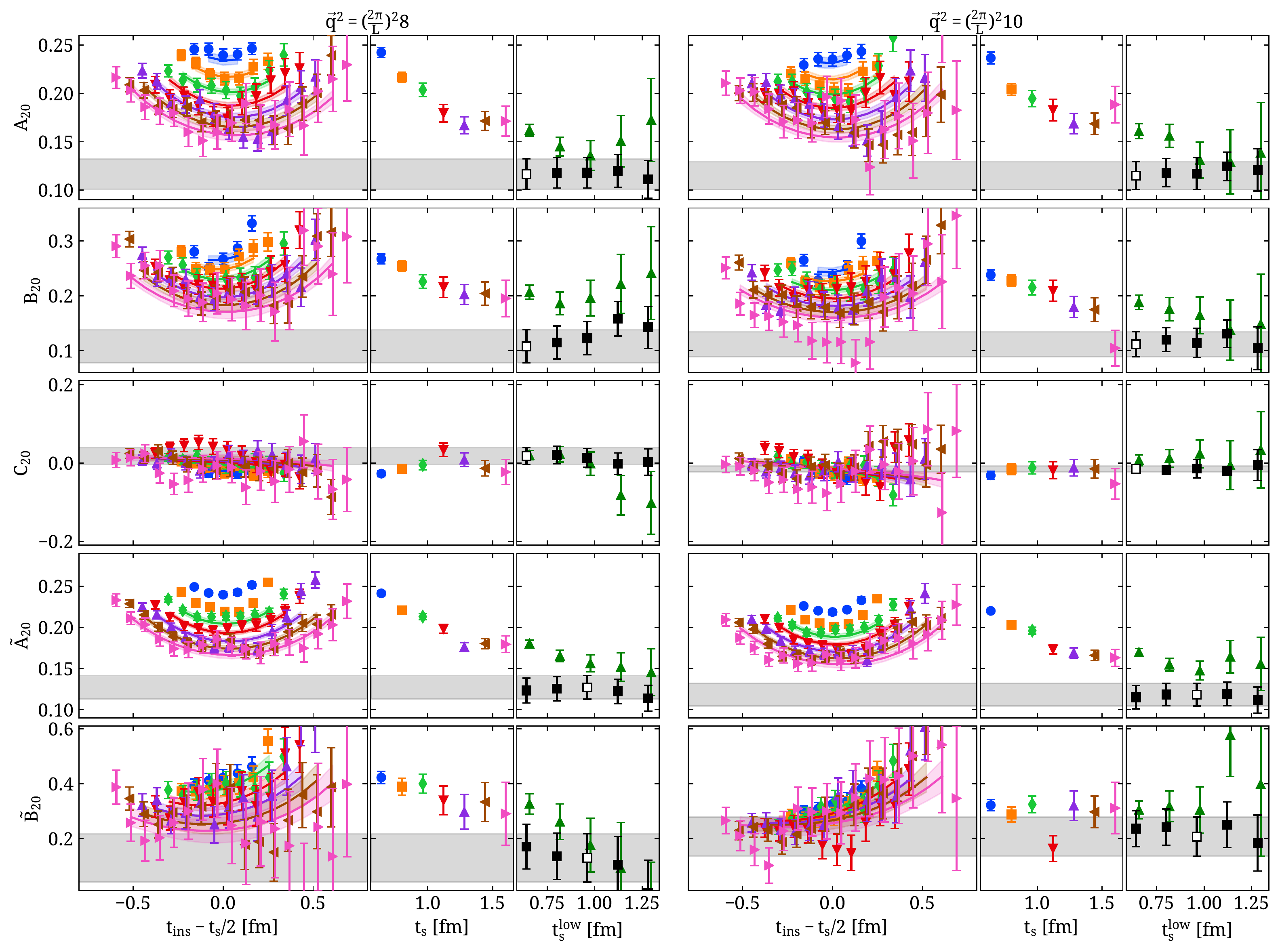}
      \caption{The same as in Fig.~\ref{fig:fits qsq 2 4} but for
        $Q^2{=}0.421$~GeV$^{2}$ (left) and
        $Q^2{=}0.514$~GeV$^{2}$ (right).}\label{fig:fits qsq 8
        10}
  \end{figure*}

To obtain the GFFs we perform, for each value of the momentum transfer
squared, a similar analysis as for the moments. This analysis is
summarized in Figs.~\ref{fig:fits qsq 2 4} and~\ref{fig:fits qsq 8 10}
for the \Nfb{} lattice for four representative values of the momentum
transfer squared. We note that for the ratios of Figs.~\ref{fig:fits
  qsq 2 4} and~\ref{fig:fits qsq 8 10} we plot, for each value of
$Q^2$ and $t_s$, the quantity: $V^\dagger \Sigma^{-1} U^\dagger R(\Gamma;
\vec{q}; t_s, t_\textrm{ins})$, where $R$ is the ratio of
Eq.~(\ref{eq:ratio}) and $U$, $\Sigma$, and $V$ are obtained from the
SVD of the kinematic matrix $\mathcal{G}$ defined in
Eq.~(\ref{eq:kin}). This is done for the purposes of presenting our
results in a similar way to that of the moments in
Fig.~\ref{fig:fits}, whereas in the analysis, to extract the result of
the plateau from the single step approach, we fit the combination
$U^\dagger R$. As in the case of the moments, we observe
non-negligible excited state effects in the ratios as we increase
$t_s$. The two-state fit results are stable as we increase
$t_s^\textrm{low}$ and the summation method converges to the two-state
fit value for $t_s^\textrm{low}\simeq$~1~fm. We therefore use the
two-state fit to extract our final values for the GFFs for all $Q^2$
using the same fit parameters as for the moments.

  \begin{figure*}
    \includegraphics[width=0.5007\linewidth]{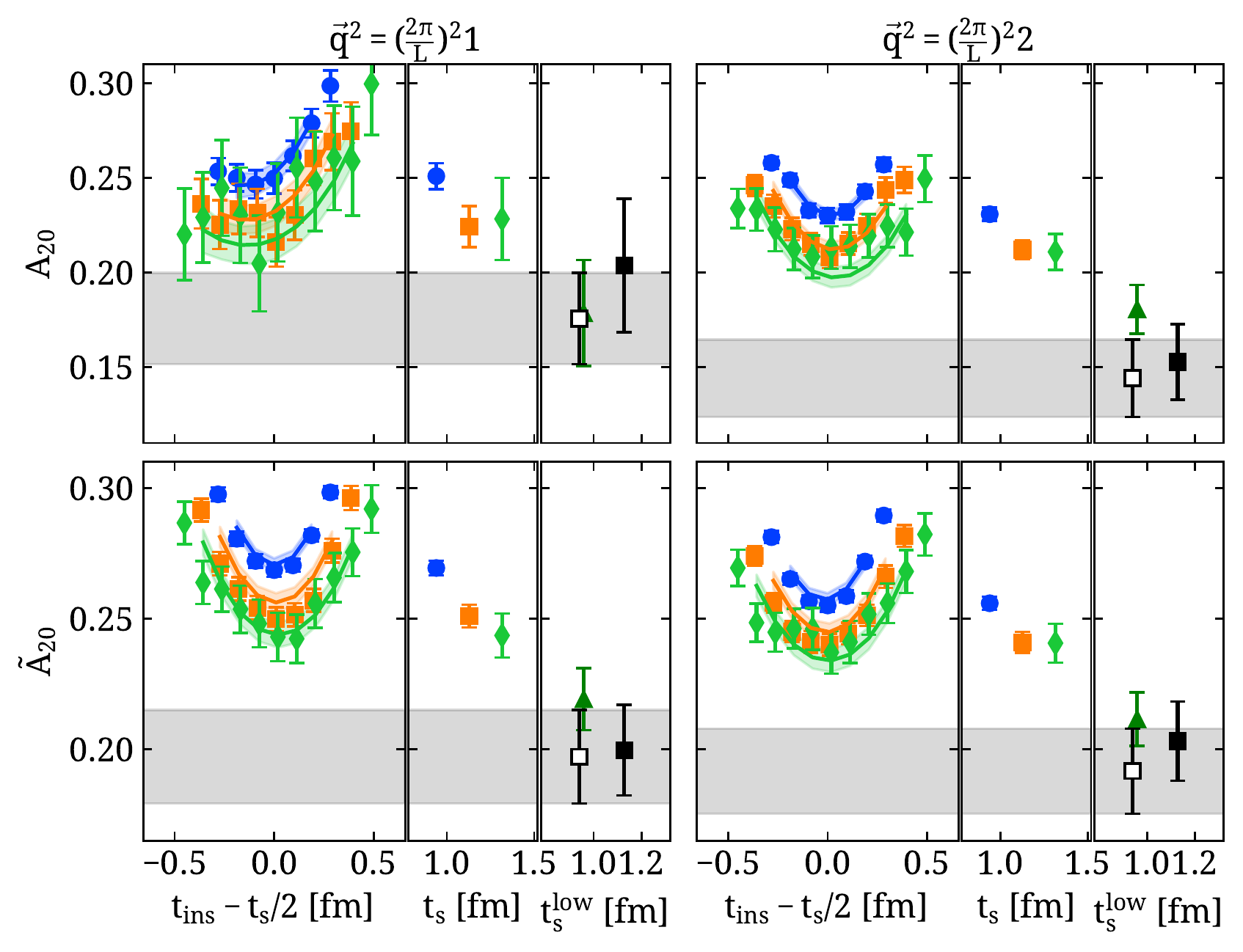}
    \includegraphics[width=0.4793\linewidth]{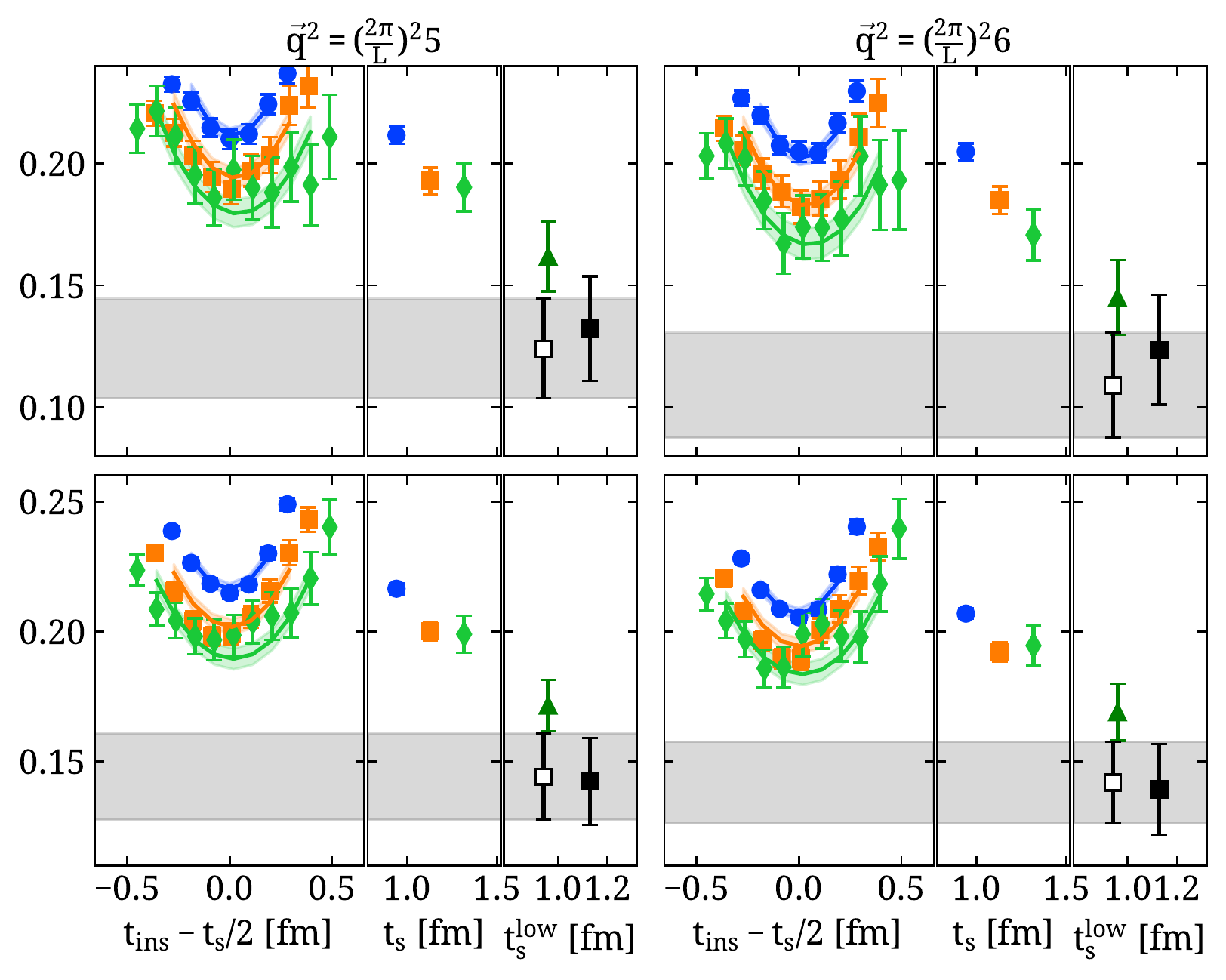}
    \caption{Results for the dominant vector (first row) and axial
      (second row) GFFs for the small \Nfa{} ensemble, cA2.09.48 for
      four representative $Q^2$ values. For each value of $Q^2$, we
      use the same notation as for the $Q^2=0$ case of
      Fig.~\ref{fig:fits cA2.09.48}.}\label{fig:fits qsq cA2.09.48}
  \end{figure*}

  \begin{figure*}
    \includegraphics[width=0.5007\linewidth]{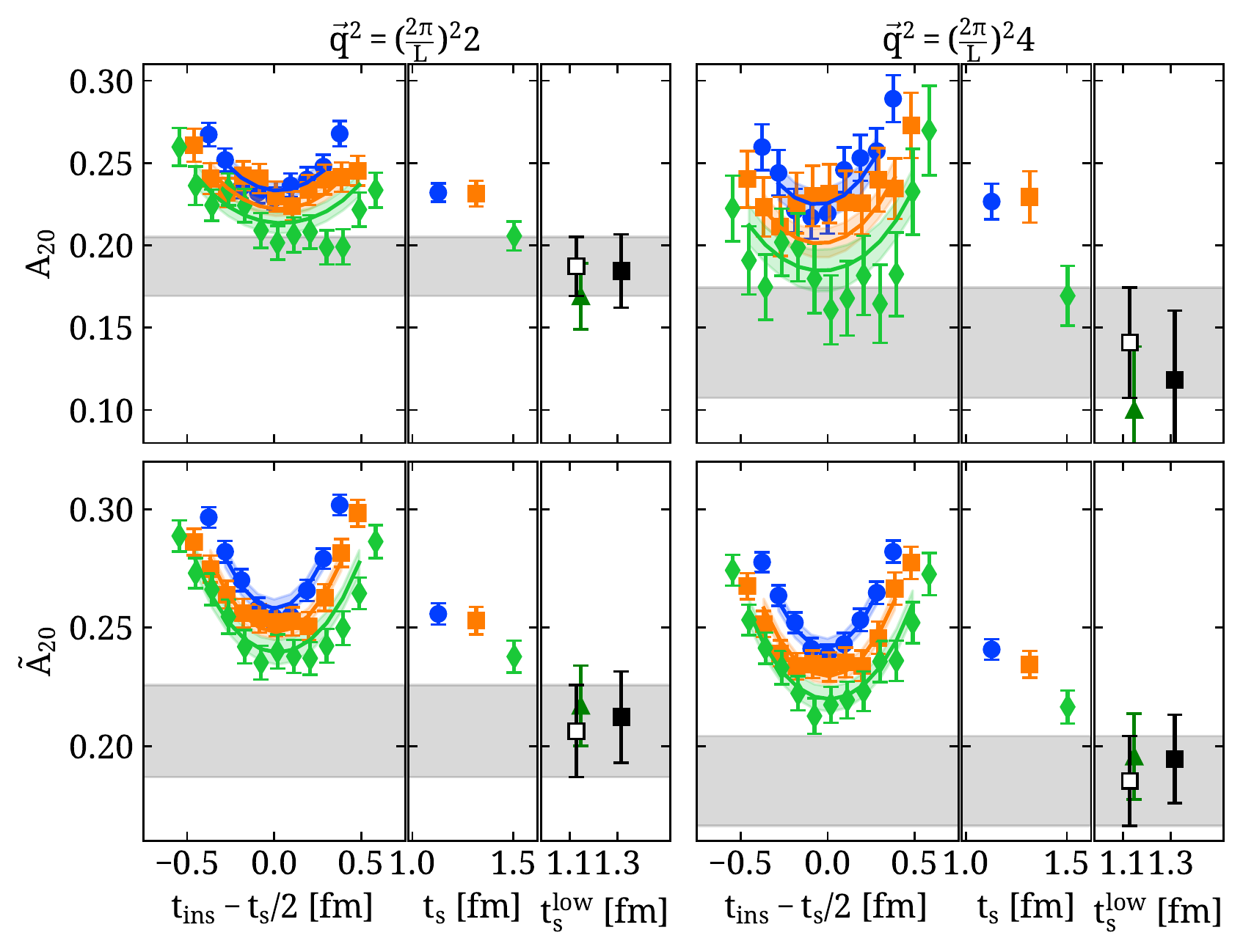}
    \includegraphics[width=0.4793\linewidth]{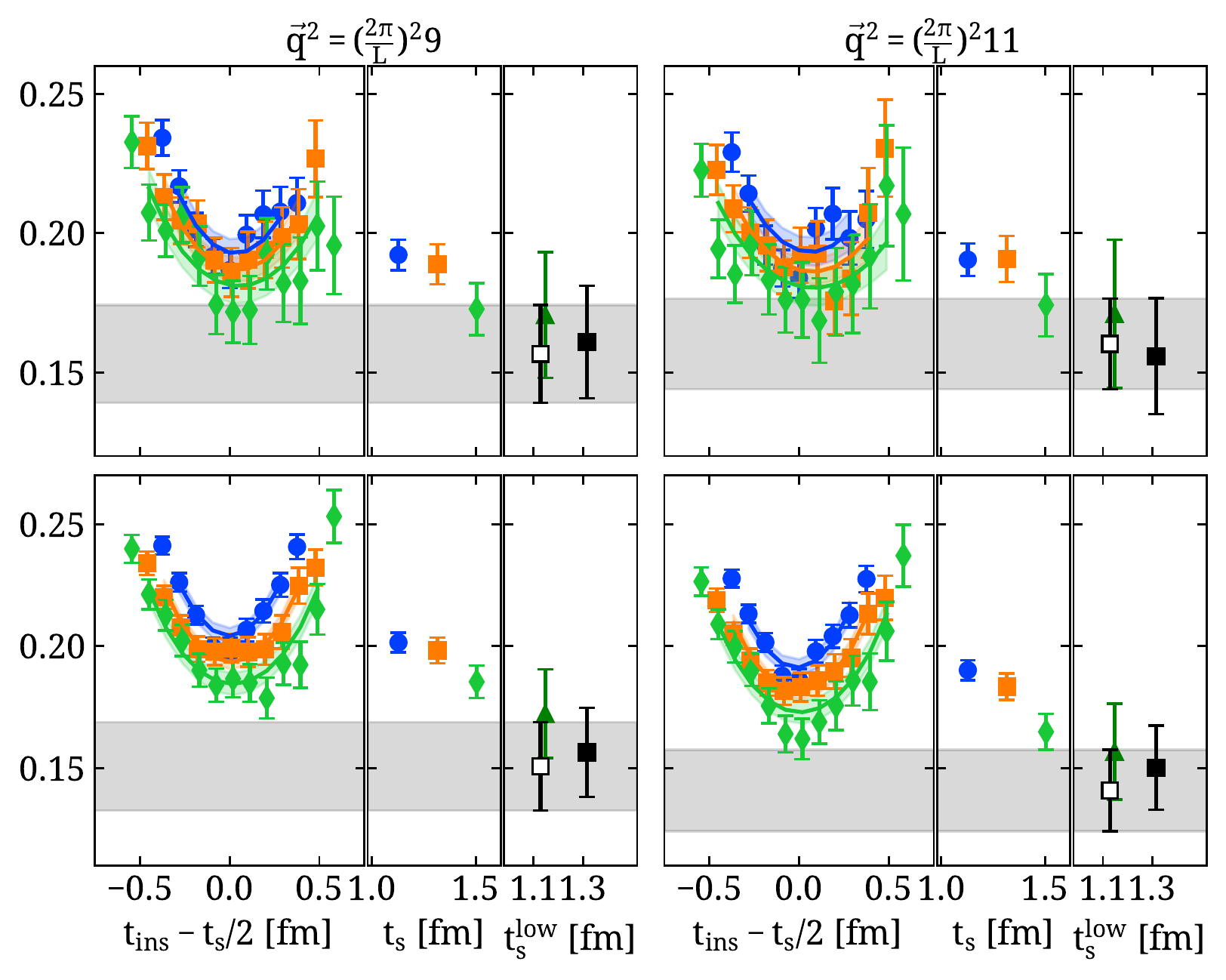}
    \caption{Results for the dominant vector (first row) and axial
      (second row) GFFs for the large \Nfa{} ensemble, cA2.09.64 for
      four representative $Q^2$ values.  For each value of $Q^2$, we
      use the same notation as for the $Q^2=0$ case of
      Fig.~\ref{fig:fits cA2.09.64}.}\label{fig:fits qsq cA2.09.64}
  \end{figure*}

The same procedure is followed for the small and large \Nfa{}
ensembles shown in Figs.~\ref{fig:fits qsq cA2.09.48}
and~\ref{fig:fits qsq cA2.09.64} respectively, in which we show the
dominant vector and axial GFFs, namely $A_{20}$ and
$\tilde{A}_{20}$. We show four representative momentum transfer values
for the two ensembles, chosen such that they are approximately equal
in physical units. Note that for extracting the vector GFFs we require
all four projectors (see Appendix~\ref{sec:appendix gff extraction}),
which means that for the small \Nfa{} ensemble we are restricted to
three sink-source separations for $Q^2{>}0$.

From Figs.~\ref{fig:fits qsq cA2.09.48} and~\ref{fig:fits qsq
  cA2.09.64}, we see that summation and two-state fit methods yield
consistent results for the $Q^2$ values shown. This is confirmed for
all $Q^2$ values, and for the subdominant vector and axial GFFs,
namely $B_{20}$, $\tilde{B}_{20}$ and $C_{20}$. 

\begin{figure*}
  \includegraphics[width=\linewidth]{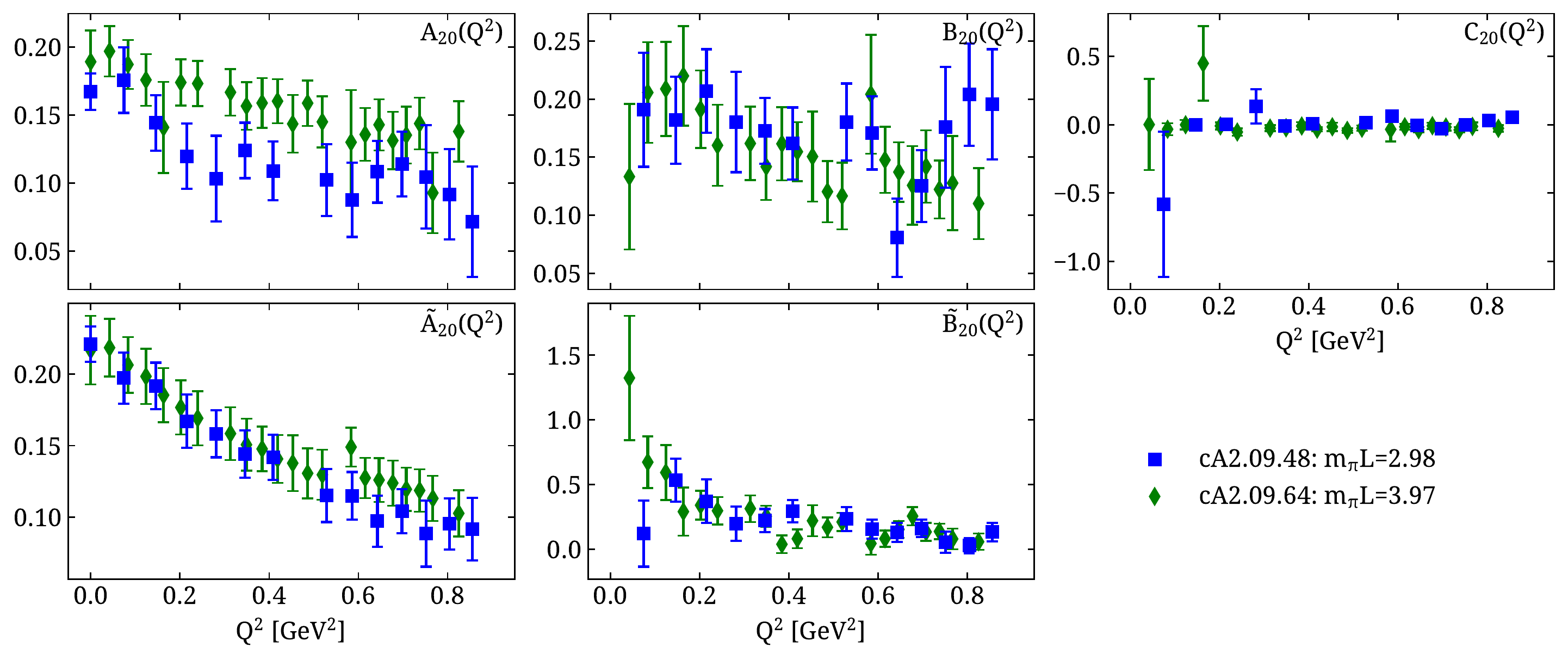}
  \caption{Comparison of the vector (top row) and axial (bottom row)
    GFFs between the two \Nfa{} ensembles cA2.09.64 (green diamonds) and
    cA2.09.48 (blue squares) which differ only in the volume, namely
    with $m_\pi L$=3.97 and 2.98 respectively. We show results
    obtained using two-state fits.}\label{fig:fve}
  \end{figure*}

The availability of the two \Nfa{} ensembles that differ only in their
volumes allows us to assess finite volume effects. A comparison
between the results obtained using these two ensembles is shown in
Fig.~\ref{fig:fve}, for all five GFFs using our final values extracted
from two-state fits. As in the case of the moments shown in
Table~\ref{table:moments}, comparing these two ensembles reveals no
finite volume effects within the achieved statistical precision. The
small discrepancies seen for $A_{20}$ at some values of the momentum
are well within the allowed statistical fluctuations. We stress that
these results are extracted taking into account correlations. Were we
to ignore the correlations among different $t_s$ the errors increase
and no disagreement is observed.

\begin{figure*}
  \includegraphics[width=\linewidth]{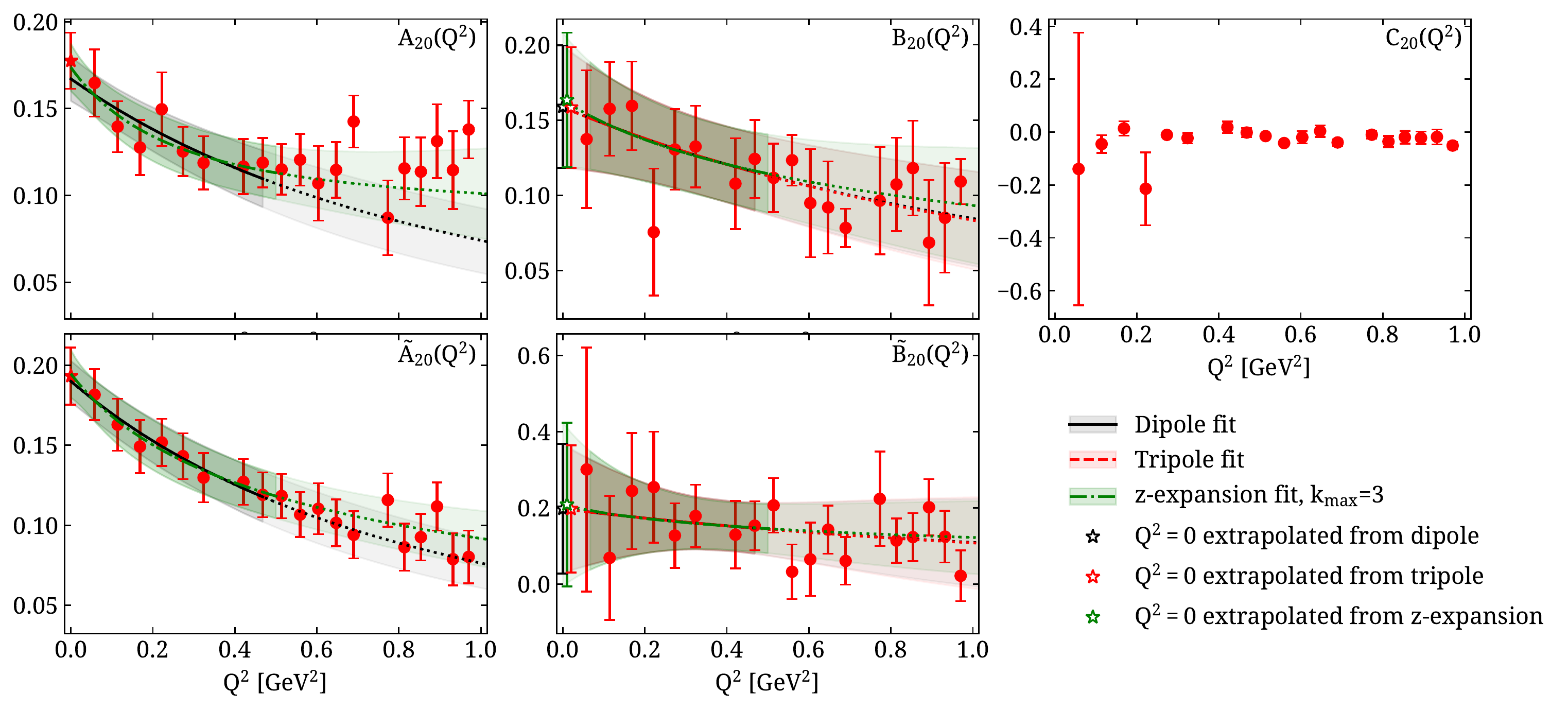}
  \caption{Results for the vector (top row) and axial (bottom row)
    GFFs for the \Nfb{} ensemble, cB211.072.64, obtained using the
    two-state fit method. Dipole (solid black curves), tripole (dashed
    red curves), and $z$-expansion (dot-dashed green curves) fits are
    shown using $Q^2\le 0.5$~GeV$^2$, while the dotted curves extend
    the fits beyond $Q^2$=0.5~GeV$^2$. For $B_{20}$ and
    $\tilde{B}_{20}$ we also show the value at $Q^2$=0 extracted from
    the dipole (open black asterisk), tripole (open red asterisk), and
    $z$-expansion (open green asterisk) fits, the latter two shifted
    slightly to improve legibility.}\label{fig:gffs}
\end{figure*}

In Fig.~\ref{fig:gffs} we show the five GFFs for the \Nfb{} ensemble
obtained from the two-state fit method. We note that $C_{20}$ is found
to be consistently zero for all $Q^2$, in agreement with previous
lattice results for this quantity~\cite{Alexandrou:2013joa,
  Bali:2018zgl}.

For the $A_{20}, B_{20}, \tilde{A}_{20}$ and $\tilde{B}_{20}$ with
nonzero signal, we perform fits to the form
\begin{equation}
  G(Q^2) = \frac{G(0)}{(1+Q^2/M^2)^n},
  \label{eq:dipole}
\end{equation}
as well as using the so-called $z$-expansion~\cite{Hill:2010yb}
\begin{equation}
  G(Q^2) = \sum_{k=0}^{k_\textrm{max}} a_k z^k,
  \label{eq:zexpansion}
\end{equation}
with:
\begin{equation}
  z = \frac{\sqrt{t_\textrm{cut} + Q^2} - \sqrt{t_\textrm{cut}}}{\sqrt{t_\textrm{cut}+Q^2} + \sqrt{t_\textrm{cut}}}.
  \label{eq:z}
\end{equation}
The dipole form, obtained by setting $n=2$ in Eq.~(\ref{eq:dipole}),
is supported by model considerations as e.g.  in the quark-soliton
model in the large $N_C$ limit for
$Q^2<1$~GeV$^2$~\cite{Goeke:2007fp}. Fitting to our data for $Q^2\le
0.5$~GeV$^2$ allowing $G(0)$ and $M$ to vary, we obtain the results
shown in Table~\ref{table:dipole fits}, where we also include the
$\chi^2$ per degrees of freedom (d.o.f) which indicates that this
Ansatz models our data well. In Fig.~\ref{fig:gffs} we show the
resulting fit to the data with the solid line. For the case of the
GFFs $B_{20}$ and $\tilde{B}_{20}$ we also consider the tripole form
by setting $n = 3$ in Eq.~(\ref{eq:dipole}). Such a form has been
shown to satisfy certain constraints in the energy and pressure
distributions inside the nucleon~\cite{Lorce:2018egm}. The resulting
tripole fit (dashed line in Fig.~\ref{fig:gffs}) is fully consistent
with the dipole yielding similar values for $B_{20}(0)$ and
$\tilde{B}_{20}(0)$ as the dipole form, as can be seen in
Table~\ref{table:dipole fits}.

The $z$-expansion provides for a model-independent Ansatz and has been
originally developed for fitting electromagnetic~\cite{Hill:2010yb}
and axial~\cite{Bhattacharya:2011ah} form factors. In
Eq.~(\ref{eq:z}), we use $t_\textrm{cut} = (4m_\pi)^2$ and
$(3m_\pi)^2$ for the vector and axial cases respectively, and fit
varying the parameters $a_k$ studying their convergence as we increase
$k_\textrm{max}$. Without loss of generality, demanding that the GFFs
are zero as $Q^2\rightarrow \infty$ constrains one parameter, which we
implement by setting
$a_{k_\textrm{max}}=-\sum_{k=0}^{k_\textrm{max}-1} a_k$. Furthermore,
in the fit we use priors for the parameters $a_k$ for
$1<k<k_\textrm{max}$. The prior width is determined as $5\max(a_0,
a_1)$, as obtained from the fit using $k_\textrm{max}${=}2. As we
increase $k_\textrm{max}$, we find that the parameters $a_0$ and $a_1$
do not change after $k_\textrm{max}=3$, for which we quote the fit
parameters in Table~\ref{table:dipole fits}. To compare with the
dipole fit, for the $z$-expansion we quote: $M=\sqrt{-8 a_0
  t_\textrm{cut}/a_1}$, which is the dipole mass that yields the same
slope as the $z$-expansion for the GFF at $Q^2=0$.

\begin{table}
  \caption{The parameters extracted from fitting $A_{20}, B_{20},
    \tilde{A}_{20}$ and $\tilde{B}_{20}$ to the dipole form ($n=2$ in
    Eq.~(\ref{eq:dipole})) and the $z$-expansion
    (Eq.~(\ref{eq:zexpansion})) and $B_{20}$ and $\tilde{B}_{20}$ to
    the tripole form ($n=3$ in Eq.~(\ref{eq:dipole})). We use $Q^2\le
    0.5$~GeV$^2$. For the $z$-expansion we show results for
    $k_\textrm{max}=3$, with $G(0)=a_0$ and $M=\sqrt{-\frac{8 a_0
        t_\textrm{cut}}{a_1}}$.}
  \label{table:dipole fits}
  \begin{tabular}{cccc}
    \hline\hline
    & $G(0)$ & $M$ [GeV] & $\chi^2$/d.o.f \\
    \hline
    & \multicolumn{3}{c}{Dipole}\\
    ${A}_{20}      $ & 0.167(13)      & 1.41(27)       & 0.4 \\
    ${B}_{20}      $ & 0.159(41)      & 1.64(72)       & 0.5 \\
    $\tilde{A}_{20}$ & 0.190(13)      & 1.32(17)       & 0.1 \\
    $\tilde{B}_{20}$ & 0.20(17)       & 1.7(2.8)       & 0.2 \\
    \hline
    & \multicolumn{3}{c}{Tripole}\\
    ${B}_{20}      $ & 0.158(40)      & 2.05(88)       & 0.5 \\
    $\tilde{B}_{20}$ & 0.20(17)       & 2.1(3.4)       & 0.2 \\
    \hline
    & \multicolumn{3}{c}{$z$-expansion ($k_\textrm{max}=3$)}\\
    ${A}_{20}      $ & 0.174(14)      & 1.03(28)       & 0.2 \\
    ${B}_{20}      $ & 0.163(45)      & 1.36(90)       & 0.4 \\
    $\tilde{A}_{20}$ & 0.195(15)      & 1.08(47)       & 0.1 \\
    $\tilde{B}_{20}$ & 0.21(21)       & 1.2(2.1)       & 0.2 \\
    \hline\hline
  \end{tabular}
\end{table}

\subsection{Comparison of results with other studies}
We compare our results with phenomenology as well as other lattice
studies with physical or near-physical pion masses.

The isovector momentum fraction $\langle x \rangle_{u-d}$ has been
extensively calculated in lattice QCD at pion masses larger than its
physical value, and a review of results can be found in
Ref.~\cite{Lin:2017snn}. Recent results include results using CLS
N\textsubscript{f}=2+1 clover improved Wilson fermions from the Mainz
group~\cite{Harris:2019bih}, as well as from two collaborations at
physical or near-physical pion mass: RQCD, using \Nfa{} clover
improved Wilson fermions~\cite{Bali:2018zgl} and
LHPC~\cite{Green:2012ud} using N\textsubscript{f}=2+1 HEX-smeared
clover improved fermions. RQCD analyzed 11 ensembles among which one
that has near-physical pion mass of 150~MeV, a lattice volume of
$64^3\times 64$ and $a=0.071$~fm. The authors analyzed three
sink-source time separations for this ensemble within $t_s\in[0.6,
  1.1]$~fm, and conclude that suppressing the excited states would
require additional separations that agrees with our findings. They,
therefore, restrict themselves to showing results using a single
separation at $15a\simeq 1.1$~fm, which is too small to control
excited states. Their value of 0.213(11)(04) is compatible with the
one we find at the similar sink-source time separation of 1.12~fm,
namely 0.232(11), that clearly overestimates the momentum fraction
extracted from larger values of $t_s$ and from the two-state fit. We
thus do not include this result in our comparison.  LHPC analyzed one
ensemble with pion mass of $m_\pi$=149~MeV, a lattice volume of
48$^3\times$48 and $a=0.116$~fm. The summation method is used to
obtain their final value from three sink-source separations with
values 0.9, 1.2, and 1.4~fm~\footnote{We note that, although this
  analysis reduced the value of the momentum fraction, performing the
  same analysis also reduced the value of the nucleon axial charge
  enlarging the discrepancy with the experimental value.}.

Results for $\langle x \rangle_{u-d}$ are shown in Fig.~\ref{fig:xq
  world} where we include the phenomenological values extracted from
global fits to PDF experimental data from Refs.~\cite{Ball:2017nwa,
  Dulat:2015mca, Harland-Lang:2014zoa, Alekhin:2017kpj,
  Accardi:2016qay, Abramowicz:2015mha}. Results from our three
ensembles are consistent with each other, indicating no detectable
lattice artifacts within their precision. Results for the \Nfb{}
ensemble are obtained using more time separations allowing for a more
rigorous assessment of exited state effects compared to the other two
ensembles. We thus take the value extracted from the \Nfb{} ensemble
to compare with phenomenology. We observe agreement with two of the
phenomenological extractions shown in Fig.~\ref{fig:xq world}, with
the remaining within 1.5$\sigma$ of our value.

\begin{figure}[h]
  \includegraphics[width=\linewidth]{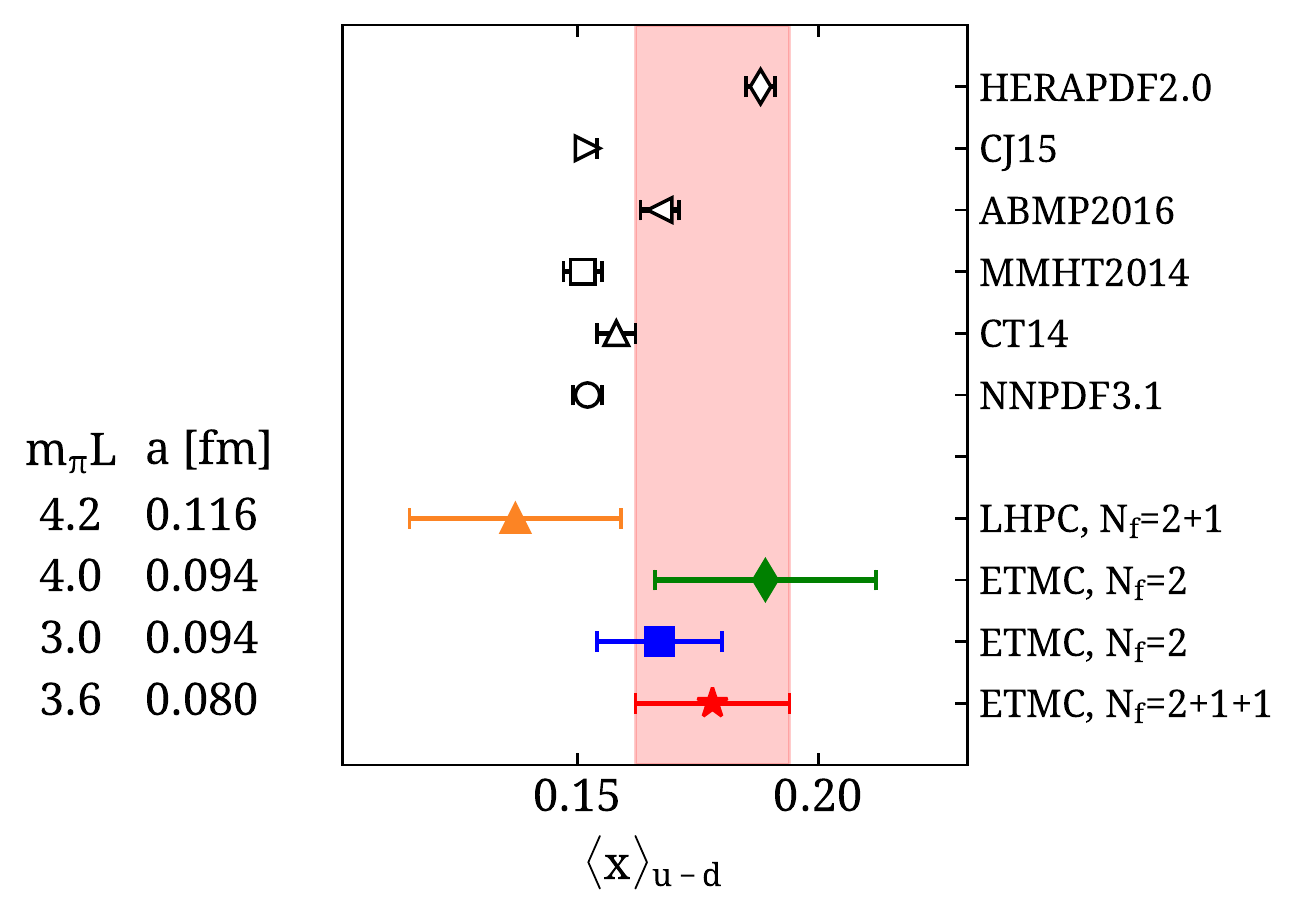}
  \caption{$\langle x \rangle_{u-d}$ from the three ensembles studied
    in this work, namely the \Nfb{} (red star), small \Nfa{} (blue
    square), and large \Nfa{} (green diamond) ensembles. We compare to
    lattice results from Ref.~\cite{Green:2012ud} (orange
    triangle). We also show results from global fits to PDF
    experimental data with the open symbols, namely
    NNPDF3.1~\cite{Ball:2017nwa} (circle), CT14~\cite{Dulat:2015mca}
    (triangle), MMHT2014~\cite{Harland-Lang:2014zoa} (square),
    ABMP2016~\cite{Alekhin:2017kpj}(left-pointing triangle),
    CJ15~\cite{Accardi:2016qay} (right-pointing triangle), and
    HERAPDF2.0~\cite{Abramowicz:2015mha} (diamonds).  }\label{fig:xq
    world}
\end{figure}

\begin{figure}[h]
  \includegraphics[width=\linewidth]{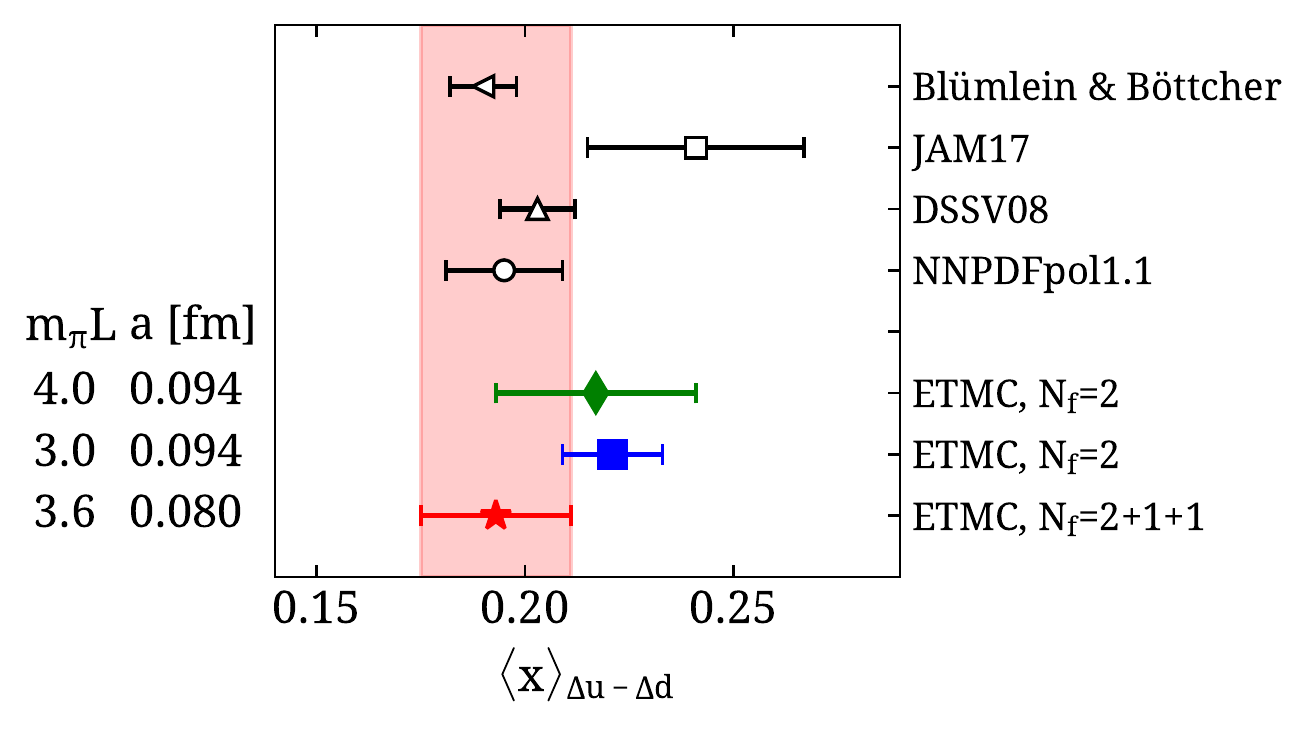}
  \caption{$\langle x \rangle_{\Delta u-\Delta d}$ from the three
    ensembles studied in this work with the notation of
    Fig.~\ref{fig:xq world}. We compare to results from global fits to
    polarized PDF experimental data with the open symbols, namely from
    Ref.~\cite{Blumlein:2010rn} (left-pointing triangle),
    NNPDFpol1.1~\cite{Nocera:2014gqa} (circle),
    DSSV08~\cite{deFlorian:2009vb} (triangle), and
    JAM17~\cite{Ethier:2017zbq} (square).  }\label{fig:xDq world}
\end{figure}

\begin{figure}[h]
  \includegraphics[width=\linewidth]{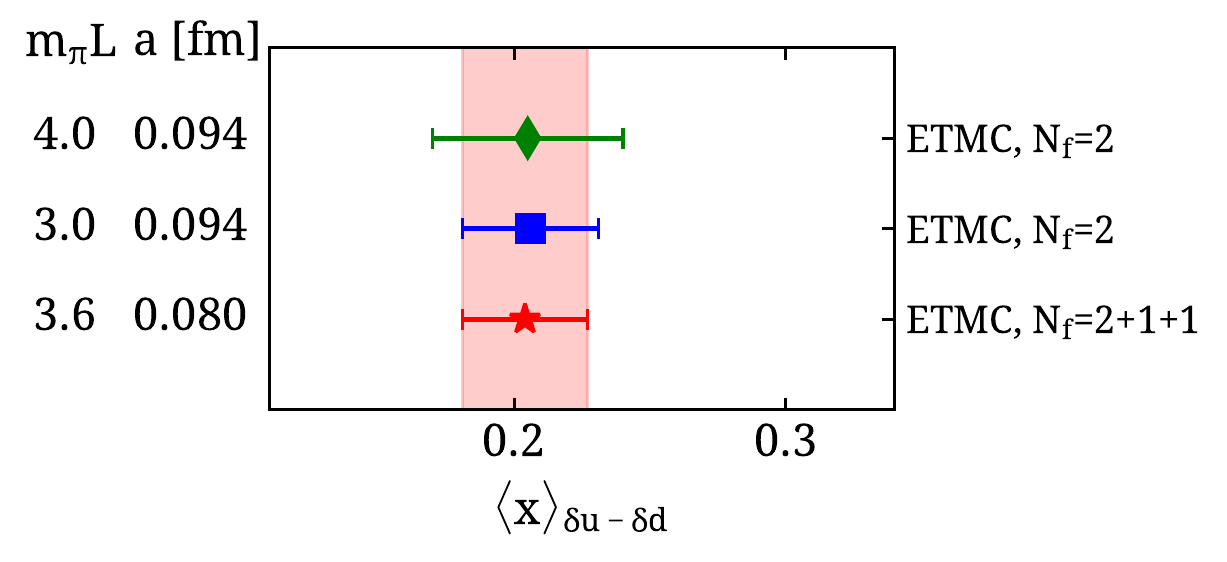}
  \caption{$\langle x \rangle_{\delta u-\delta d}$ from the three
    ensembles studied in this work with the notation of
    Fig.~\ref{fig:xq world}.}\label{fig:xtq world}
\end{figure}

For $\langle x \rangle_{\Delta u-\Delta d}$ our results are compared
in Fig.~\ref{fig:xDq world} to phenomenological results from
Refs.~\cite{Ethier:2017zbq,Nocera:2014gqa,Blumlein:2010rn,deFlorian:2009vb}.
As can be seen, our value is in good agreement with these
phenomenological determinations and in particular with the value found
in Ref.~\cite{Blumlein:2010rn}. The results for $\langle x
\rangle_{\delta u-\delta d}$ are shown in Fig.~\ref{fig:xtq world}. No
phenomenological nor other lattice QCD results at the physical point
are available for the tensor moment and thus the current work provides
a valuable prediction. We note that for the helicity and tensor
moments only three sink-source time separations are available in the
case of the two \Nfa{} ensembles. This restricts the two-state
analysis and thus we consider the result of the \Nfb{} ensemble as the
most reliable.  As already mentioned, the two \Nfa{} ensembles show no
detectable volume dependence for these quantities indicating that
volume effects are negligible as compared to the current accuracy
obtained from the analysis using the \Nfb{} ensemble.

\section{Conclusions}
\label{sec:conclusions}
The isovector momentum fraction, helicity moment, and transversity of
the nucleon are extracted using lattice QCD simulations produced with
physical values of the quark masses.  For the \Nfb{} ensemble
cB211.072.64, seven sink-source separations are analyzed from 0.6~fm
to 1.6~fm, allowing for the most thorough study of excited states to
date for these quantities directly at the physical pion mass.  The
isovector unpolarized and helicity GFFs are also extracted for the
first time directly at the physical point. The study reveals that both
for $Q^2=0$ as well as for $Q^2>0$, the convergence of these
quantities to the ground state is slow. For values of the sink-source
time separation $t_s<2$~fm a two-state fit analysis yields stable
results and agrees with the values extracted from the summation method
when including separations larger than $\sim$1~fm. We therefore, take
the results from the two-state fit when confirmed with the summation
method as our final values.  The results for the GFFs are provided in
Tables~\ref{table:gffs cB211.072.64},~\ref{table:gffs cA2.09.48},
and~\ref{table:gffs cA2.09.64} of Appendix~\ref{sec:appendix results}
and in Table~\ref{table:moments} for the moments. For the case of the
small \Nfa{} ensemble, the current results for the moments are an
update to those of Ref.~\cite{Abdel-Rehim:2015owa}, which are included
for comparison in Appendix~\ref{sec:appendix 2015 comparison}. The
\Nfb{} ensemble includes dynamical strange and charm quarks in
addition to the light quarks thus providing a full description of the
QCD vacuum. In addition, the seven sink-source separations are
analyzed to high accuracy allowing for a robust analysis of excited
states. We thus consider the results extracted from the \Nfb{}
ensemble as the best prediction of these quantities. We thus quote as
our final results the values obtained from the analysis of the \Nfb{}
ensemble. We find for the moments:
\begin{align}
  \langle x \rangle_{u-d}             &= 0.178(16),  \nonumber\\
  \langle x \rangle_{\Delta u-\Delta d} &= 0.193(18), \nonumber\\
  \langle x \rangle_{\delta u-\delta d} &= 0.204(23),
\end{align}
where we quote the values extracted directly from the nucleon matrix
element at zero momentum.  The values for the unpolarized and helicity
moments agree with a subset of the phenomenological results.  The
helicity and transversity moments $\langle x\rangle_{\Delta q}$ and
$\langle x\rangle_{\delta q}$, are shown to be related to longitudinal
and transverse spin-orbit correlations,
respectively~\cite{Lorce:2014mxa,Bhoonah:2017olu} and are interpreted
as the parity and chiral partners of Ji's relation for angular
momentum.

Fits of the GFFs yield the results provided in Table~\ref{table:dipole
  fits}. From these fits we obtain $B_{20}(0)$, which is related to the
proton spin via Ji's sum rule~\cite{Ji:1996ek}.  Using the values for
\Nfb{} obtained by fits to the dipole form, we obtain:
\begin{equation}
  J^{u-d}=\frac{1}{2}[A^{u-d}_{20}(0) + B^{u-d}_{20}(0)] = 0.168(22)(02)
\end{equation}
for the isovector contribution of the up and down quarks to the
proton spin, where the first error is statistical and the second a
systematic obtained as the difference in $B_{20}$ between the
extraction using the dipole form and the $z$-expansion.

A next step in this study will be the inclusion of disconnected
contributions in order to calculate the isoscalar and gluonic
quantities.  This would allow for complete flavor decomposition of the
GFFs and for calculating the spin and momentum carried by quarks and
gluons in the proton.

\section*{ACKNOWLEDGMENTS}
We thank all members of the Extended Twisted Mass Collaboration for a
very constructive and enjoyable collaboration. M.C. acknowledges
financial support by the U.S. National Science Foundation under Grant
No. PHY-1714407 and C.U. by the DFG as a project in the Sino-German
CRC110. Computational resources from Extreme Science and Engineering
Discovery Environment (XSEDE) were used, which is supported by
National Science Foundation Grant No. TG-PHY170022. C.L.
acknowledges support from the U.S. Department of Energy, Office of
Science, Office of Nuclear Physics, Contract No. DE-AC02-06CH11357.
This project has received funding from the Horizon 2020 research and
innovation program of the European Commission under the Marie
Sk\l{}odowska-Curie Grant Agreement No. 765048.  The authors
gratefully acknowledge the Gauss Centre for Supercomputing
e.V. (www.gauss-centre.eu) for funding this project by providing
computing time on the GCS Supercomputer SuperMUC at Leibniz
Supercomputing Centre (www.lrz.de) under Project pr74yo and through
the John von Neumann Institute for Computing (NIC) on the GCS
Supercomputers JUQUEEN~\cite{juqueen}, JURECA~\cite{jureca} and
JUWELS~\cite{JUWELS} at Jülich Supercomputing Centre (JSC), under
Projects ECY00 and HCH02. This work was supported by a grant from the
Swiss National Supercomputing Centre (CSCS) under Project ID s702. We
thank the staff of CSCS for access to the computational resources and
for their constant support. This research uses resources of Temple
University, supported in part by the National Science Foundation
(Grant No. 1625061) and by the U.S. Army Research Laboratory (Contract
No. W911NF-16-2-0189).

\bibliography{refs.bib}


\appendix
\widetext

\section{Expressions for generalized form factors}
\label{sec:appendix gff extraction}
The following expressions are provided in Euclidean space. We suppress
the $Q^2=-q^2$ argument of the generalized form factors, $E_N$ is the
nucleon energy for three-momentum $\vec{q}$, the kinematic factor
$\mathcal{K} = \sqrt{2m_N^2/[E_N(E_N+m_N)]}$ and Latin indices ($k$,
$n$, and $j$) take values 1, 2, and 3 with $k\neq j$ while $\rho$
takes values 1, 2, 3, and 4.

\subsection{Vector operator}
\begin{align}
  \Pi_V^{00}(\Gamma^0, \vec q) =& A_{20}\,\mathcal{K}\,\left(-\frac{3\,E_N}{8} - \frac{E_N^2}{4\,m_N} -
  \,\frac{m_N}{8} \right) +
  B_{20}\,\mathcal{K}\,\left( -\,\frac{E_N }{8} +
  \,\frac{E_N^3}{8\,m_N^2} + \frac{E_N^2}{16\,m_N} - \frac{m_N}{16}
  \right)\nonumber\\
  +&   C_{20}\,\mathcal{K}\,\left(\,\frac{E_N}{2} - \frac{E_N^3}{2\,m_N^2} +
  \frac{E_N^2}{4\,m_N} - \frac{m_N}{4} \right) ,\\
  \Pi_V^{00}(\Gamma^n, \vec q) =& 0 ,\\
  \Pi_V^{kk}(\Gamma^0, \vec q) =&  A_{20}\,\mathcal{K}\,\left(
  \frac{E_N}{8} + \frac{m_N}{8} + \frac{q_k^2}{4\,m_N} \right)  +
  B_{20}\,\mathcal{K}\,\left( -\frac{E_N^2}{16\,m_N} + \frac{m_N}{16} - \frac{q_k^2\,E_N}{8\,m_N^2} +
  \frac{q_k^2}{8\,m_N} \right)  \nonumber \\
  +&  C_{20}\,\mathcal{K}\,\left( -\frac{E_N^2}{4\,m_N} + \frac{m_N}{4} + \frac{q_k^2\,E_N}{2\,m_N^2} +
  \frac{q_k^2}{2\,m_N} \right)  ,\\
  \Pi_V^{kk}(\Gamma^n, \vec q) =&  A_{20}\,\mathcal{K}\,\left(-i\,\frac{\epsilon_{k\,n\,0\,\rho}\,
    q_k\,q_\rho }{4\,m_N}\right) +
  B_{20}\,\mathcal{K}\,\left( -i\,\frac{\epsilon_{k\,n\,0\,\rho}\,
    q_k\,q_\rho }{4\,m_N}\right),\\
  \Pi_V^{k0}(\Gamma^0, \vec q) =&  A_{20}\,\mathcal{K}\,\left(-i\,\frac{q_k}{4} -i\,\frac{q_k\,E_N}{4\,m_N} \right)+
  B_{20}\,\mathcal{K}\,\left(-i\,\frac{q_k}{8} +i\, \frac{q_k\,E_N^2}{8\,m_N^2}  \right)+
  C_{20}\,\mathcal{K}\,\left(i\,\frac{q_k}{2}-i\, \frac{q_k\,E_N^2}{2\,m_N^2}  \right),\\
  \Pi_V^{k 0}(\Gamma^n, \vec q) =&  A_{20}\,\mathcal{K}\,\left(\,-\epsilon_{k\,n\,0\,\rho}\,\left(\frac{
    q_\rho}{8} +\frac{q_\rho\,E_N}{8\,m_N}  \right) \right) +
  B_{20}\,\mathcal{K}\,\left(\,-\epsilon_{k\,n\,0\,\rho}\,\left(\frac{q_\rho}{8} +
  \frac{q_\rho\,E_N}
       {8\,m_N}   \right)\right) 
\end{align}
\begin{align}
       \Pi_V^{kj}(\Gamma^0, \vec q) =&    A_{20}\,\mathcal{K}\,\frac{q_k\,q_j}
          {4\,m_N} + B_{20}\,\mathcal{K}\,\left(-\frac{
            q_k\,q_j \,E_N}{8\,m_N^2} +
          \frac{q_k\,q_j}{8\,m_N} \right)  +
          C_{20}\,\mathcal{K}\,\left( \frac{q_k\,
            q_j\,E_N}{2\,m_N^2} +
          \frac{q_k\,q_j}{2\,m_N} \right) ,\\
          \Pi_V^{kj}(\Gamma^n, \vec q) =&  A_{20}\,\mathcal{K}\,\left(-i\,\frac{\epsilon_{k\,n\,0\,\rho}\,
            q_j\,q_\rho }{8\,m_N}-i\,\frac{\epsilon_{j\,n\,0\,\rho}\,
            q_k\,q_\rho }{8\,m_N}\right) +
          B_{20}\,\mathcal{K}\,\left(-i\,\frac{\epsilon_{k\,n\,0\,\rho}\,
            q_j\,q_\rho }{8\,m_N}-i\,\frac{\epsilon_{j\,n\,0\,\rho}\,
            q_k\,q_\rho }{8\,m_N}\right).
\end{align}
\subsection{Axial operator}

\begin{align}
  \Pi_A^{\mu\nu}(\Gamma^0, \vec q) =&   0,\\
  \Pi_A^{k0}(\Gamma^n, \vec q) =&  \tilde{A}_{20}\,\mathcal{K}\,\left(\,-i\, \delta_{n\,k}\,
  \left( \frac{E_N}{4} + \frac{E_N^2}{8\,m_N} + \frac{m_N}{8} \right)  -
  i\,\frac{q_k\,q_n}{8\,m_N} \right) +
  \tilde{B}_{20}\,\mathcal{K}\,\left(i\,\frac{q_k\,q_n \,E_N}{8\,m_N^2}  \right),\\
  \Pi_A^{kj}(\Gamma^n, \vec q) =& \tilde{A}_{20}\,\mathcal{K}\,\left(\delta_{n\,j}\,
  \left(\frac{q_k }{8} +
  \frac{q_k\,E_N}{8\,m_N} \right)  
  + \delta_{n\,k}\,\left(\frac{q_j }
  {8} +\frac{q_j\,E_N}{8\,m_N} \right)  \right) +
  \tilde{B}_{20}\,\mathcal{K}\,\left(-\frac{q_k\,q_j\,q_n}{8\,m_N^2} \right).
\end{align}

\section{Effective energies and dispersion relation}
\label{sec:appendix dispersion}
\begin{figure}
    \includegraphics[width=0.475\linewidth]{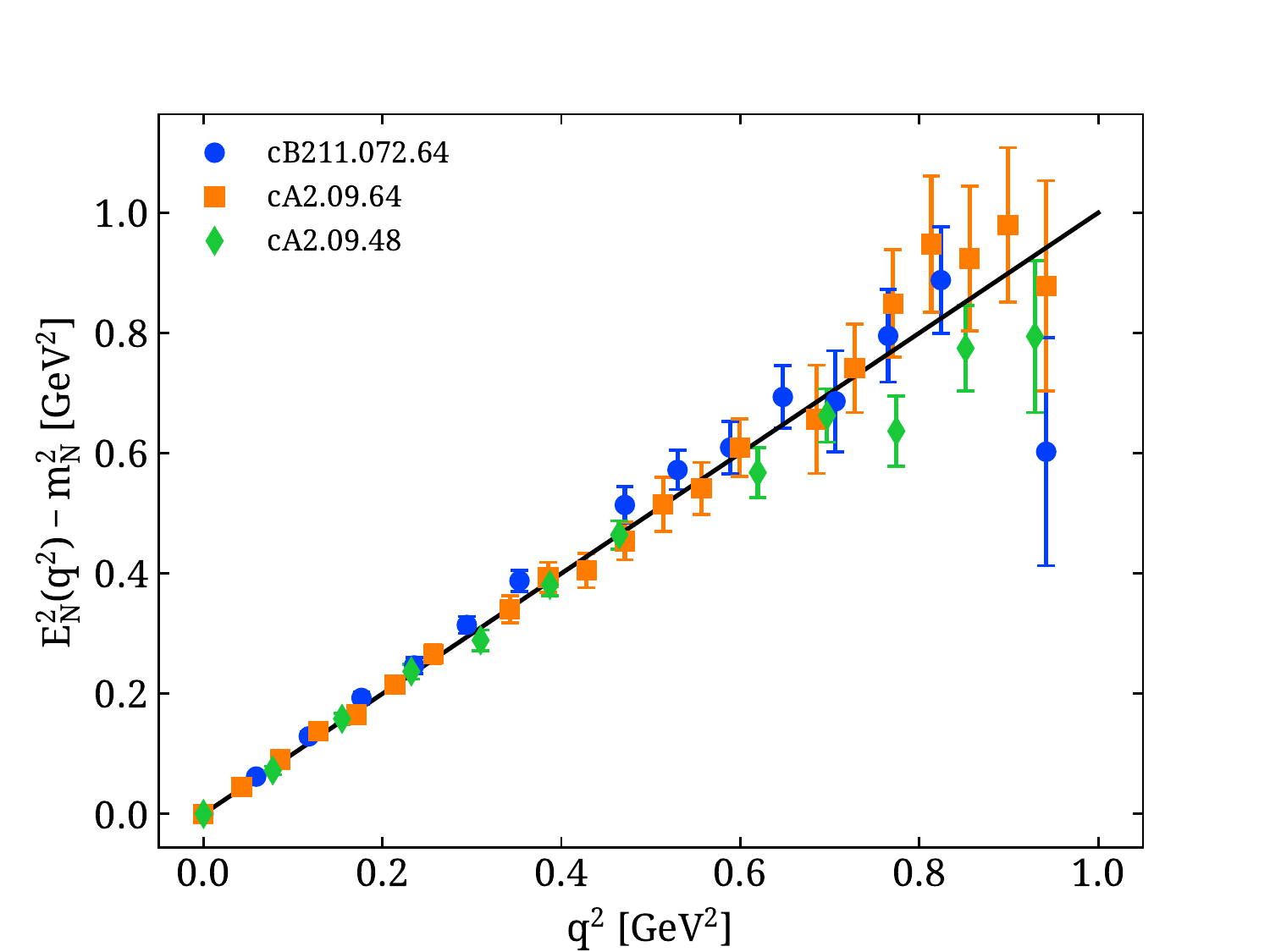}
    \caption{The difference between squared effective energy
      and squared effective mass obtained for the \Nfb{} (blue
      circles), the large \Nfa{} (orange squares), and the small
      \Nfa{} (green diamonds) ensembles as a function of the spatial
      momentum squared. The black line is the continuum dispersion
      relation $E^2_N(q^2) - m_N^2 = q^2$.}
    \label{fig:disp}
  \end{figure}

In Fig.~\ref{fig:disp} we plot for the three ensembles used in this
work the energies obtained by fits to the effective energy as a
function of the two-point function momentum. As can be seen, the
effective energies obtained are consistent with the continuum
dispersion relation $E_N(\vec{q})=\sqrt{\vec{q}^2+m_N^2}$, where
$\vec{q}=\frac{2\pi}{L}\vec{n}$ with $\vec{n}$ a lattice vector with
components $n_i\in (-\frac{L}{2a}, \frac{L}{2a}]$.
  
\section{Tables of Results}
\label{sec:appendix results}
Results for the GFFs are provided for $A_{20}(Q^2)$, $B_{20}(Q^2)$,
$\tilde{A}_{20}(Q^2)$, and $\tilde{B}_{20}(Q^2)$ using the two-state
fit method as explained in the text. We do not provide $C_{20}$ which
is found to be consistent with zero. We provide results for ensemble
cB211.072.64 in Table~\ref{table:gffs cB211.072.64}, for cA2.09.48 in
Table~\ref{table:gffs cA2.09.48}, and for cA2.09.64 in
Table~\ref{table:gffs cA2.09.64}.
\begin{table}[!h]
  \caption{GFFs for ensemble cB211.072.64 obtained using the two-state
    fit method. We do not provide $C_{20}$ which is found to be
    consistent with zero.}\label{table:gffs cB211.072.64}
  \begin{tabular}{ccccc}
    \hline\hline
    $Q^2$ [GeV$^2$] & $A_{20}(Q^2)$ & $B_{20}(Q^2)$ & $\tilde{A}_{20}(Q^2)$ & $\tilde{B}_{20}(Q^2)$ \\
    \hline
 0.000 & 0.178(16)  &            & 0.193(18)  &            \\
 0.058 & 0.165(19)  & 0.137(46)  & 0.182(16)  & 0.30(32)   \\
 0.114 & 0.140(15)  & 0.157(31)  & 0.163(16)  & 0.07(16)   \\
 0.169 & 0.128(16)  & 0.159(29)  & 0.149(17)  & 0.24(15)   \\
 0.222 & 0.150(21)  & 0.075(42)  & 0.152(15)  & 0.25(15)   \\
 0.273 & 0.125(14)  & 0.130(27)  & 0.143(14)  & 0.127(85)  \\
 0.324 & 0.119(15)  & 0.132(27)  & 0.130(15)  & 0.179(82)  \\
 0.421 & 0.117(16)  & 0.108(30)  & 0.127(14)  & 0.130(89)  \\
 0.468 & 0.119(14)  & 0.124(26)  & 0.119(14)  & 0.153(64)  \\
 0.514 & 0.115(15)  & 0.112(23)  & 0.118(14)  & 0.207(72)  \\
 0.559 & 0.121(15)  & 0.123(17)  & 0.107(14)  & 0.032(72)  \\
 0.603 & 0.107(22)  & 0.095(36)  & 0.110(16)  & 0.065(97)  \\
 0.647 & 0.115(16)  & 0.092(31)  & 0.102(15)  & 0.143(63)  \\
 0.690 & 0.143(15)  & 0.078(13)  & 0.094(15)  & 0.061(63)  \\
 0.773 & 0.087(21)  & 0.096(36)  & 0.116(17)  & 0.22(12)   \\
 0.814 & 0.116(18)  & 0.107(31)  & 0.086(15)  & 0.114(58)  \\
 0.854 & 0.114(20)  & 0.118(32)  & 0.093(15)  & 0.123(63)  \\
 0.893 & 0.131(21)  & 0.069(42)  & 0.112(15)  & 0.202(74)  \\
 0.932 & 0.115(22)  & 0.085(36)  & 0.079(16)  & 0.125(68)  \\
 0.970 & 0.138(17)  & 0.109(15)  & 0.080(16)  & 0.022(67)  \\
    \hline\hline
  \end{tabular}
\end{table}

\begin{table}[!h]
  \caption{GFFs for ensemble cA2.09.48 obtained using the two-state
    fit method. We do not provide $C_{20}$ which is found to be
    consistent with zero.}\label{table:gffs cA2.09.48}
  \begin{tabular}{ccccc}
    \hline\hline
    $Q^2$ [GeV$^2$] & $A_{20}(Q^2)$ & $B_{20}(Q^2)$ & $\tilde{A}_{20}(Q^2)$ & $\tilde{B}_{20}(Q^2)$ \\
 0.000 & 0.167(13)  &            & 0.221(12)  &            \\
 0.075 & 0.176(24)  & 0.191(49)  & 0.197(18)  & 0.12(26)   \\
 0.146 & 0.144(20)  & 0.182(38)  & 0.192(16)  & 0.53(17)   \\
 0.215 & 0.120(24)  & 0.207(36)  & 0.167(19)  & 0.37(17)   \\
 0.282 & 0.103(31)  & 0.180(43)  & 0.158(16)  & 0.20(13)   \\
 0.346 & 0.124(20)  & 0.173(28)  & 0.144(17)  & 0.220(90)  \\
 0.409 & 0.109(22)  & 0.162(31)  & 0.142(16)  & 0.294(86)  \\
 0.529 & 0.102(26)  & 0.180(33)  & 0.115(18)  & 0.233(91)  \\
 0.586 & 0.088(27)  & 0.171(32)  & 0.115(17)  & 0.154(73)  \\
 0.643 & 0.108(23)  & 0.081(34)  & 0.097(18)  & 0.130(74)  \\
 0.698 & 0.114(24)  & 0.125(31)  & 0.104(15)  & 0.160(68)  \\
 0.751 & 0.105(38)  & 0.176(52)  & 0.089(23)  & 0.053(82)  \\
 0.804 & 0.092(33)  & 0.204(44)  & 0.095(18)  & 0.026(59)  \\
 0.855 & 0.072(41)  & 0.196(47)  & 0.092(22)  & 0.132(71)  \\
    \hline\hline
  \end{tabular}
\end{table}

\begin{table}[!h]
  \caption{GFFs for ensemble cA2.09.64 obtained using the two-state
    fit method. We do not provide $C_{20}$ which is found to be
    consistent with zero.}\label{table:gffs cA2.09.64}
  \begin{tabular}{ccccc}
    \hline\hline
    $Q^2$ [GeV$^2$] & $A_{20}(Q^2)$ & $B_{20}(Q^2)$ & $\tilde{A}_{20}(Q^2)$ & $\tilde{B}_{20}(Q^2)$ \\
0     & 0.189(23)  &           & 0.217(24) &   \\
0.042 & 0.197(18)  & 0.133(63) & 0.218(20) & 1.32(48)   \\
0.083 & 0.187(18)  & 0.206(43) & 0.206(19) & 0.67(20)   \\
0.123 & 0.176(19)  & 0.209(41) & 0.198(19) & 0.59(21)   \\
0.163 & 0.141(34)  & 0.220(43) & 0.185(19) & 0.29(19)   \\
0.201 & 0.174(17)  & 0.191(33) & 0.177(19) & 0.34(11)   \\
0.239 & 0.173(17)  & 0.160(35) & 0.169(19) & 0.30(11)   \\
0.312 & 0.167(17)  & 0.162(32) & 0.158(18) & 0.31(10)   \\
0.348 & 0.157(18)  & 0.142(29) & 0.151(18) & 0.254(83)  \\
0.383 & 0.159(18)  & 0.162(32) & 0.148(16) & 0.038(69)  \\
0.417 & 0.160(16)  & 0.155(25) & 0.141(17) & 0.081(72)  \\
0.451 & 0.144(21)  & 0.151(39) & 0.138(19) & 0.22(12)   \\
0.484 & 0.159(17)  & 0.120(26) & 0.131(18) & 0.169(77)  \\
0.517 & 0.145(19)  & 0.117(29) & 0.130(17) & 0.210(71)  \\
0.581 & 0.130(38)  & 0.204(51) & 0.149(14) & 0.044(87)  \\
0.612 & 0.136(19)  & 0.148(28) & 0.127(14) & 0.081(63)  \\
0.643 & 0.143(19)  & 0.137(26) & 0.126(15) & 0.153(64)  \\
0.674 & 0.131(21)  & 0.126(34) & 0.124(16) & 0.255(70)  \\
0.704 & 0.135(21)  & 0.142(31) & 0.120(15) & 0.133(70)  \\
0.734 & 0.144(19)  & 0.122(25) & 0.119(15) & 0.137(60)  \\
0.763 & 0.093(30)  & 0.128(40) & 0.113(16) & 0.079(80)  \\
0.821 & 0.138(22)  & 0.110(31) & 0.103(16) & 0.057(63)  \\
\hline\hline
  \end{tabular}
\end{table}

\section{Comparison with previous results}
\label{sec:appendix 2015 comparison}
\begin{figure}[!h]
  \begin{center}
    \includegraphics[width=0.8\linewidth]{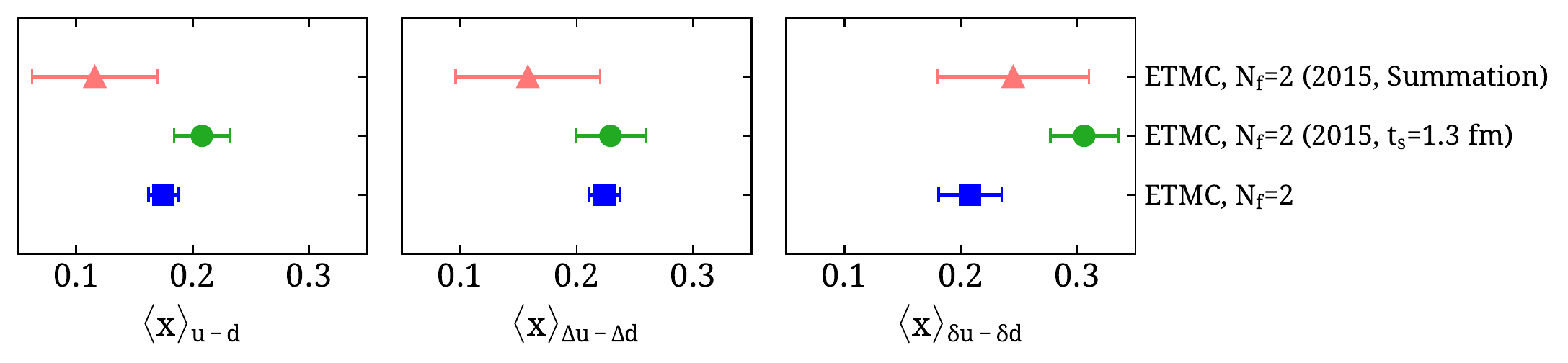}
  \end{center}
  \caption{Results for the moments obtained on the cA2.09.48 ensemble
    as obtained in this work (blue squares) compared to previous
    results as presented in Ref.~\cite{Abdel-Rehim:2015owa} using the
    plateau method at the largest sink-source separation available at
    that time ($t_s=1.3$~fm, green circles) and the summation method
    (red triangles).}
  \label{fig:2015 comparison}
\end{figure}

The results presented here for the small \Nfa{} ensemble, cA2.09.48,
are an update to those first presented in
Ref.~\cite{Abdel-Rehim:2015owa}. For completeness, in
Fig.~\ref{fig:2015 comparison} we compare the current results with
those obtained in Ref.~\cite{Abdel-Rehim:2015owa}.

\end{document}